\documentclass[a4paper,12pt]{article}
\pdfoutput=1 

\usepackage[T1]{fontenc} 

\usepackage{draft}
\usepackage{putex} 
\usepackage{soul}

\usepackage{cite}

\usepackage[
      colorlinks=true,
      linkcolor=black,
      urlcolor=blue,
      filecolor=black,
      citecolor=red,
      ]{hyperref}
      
\usepackage{graphicx}
\usepackage{xcolor}
\usepackage{amsfonts,amsmath,amssymb,bm}
\usepackage{verbatim}
\usepackage{array}
\usepackage{mathrsfs}
\usepackage{mathtools}
\usepackage{yfonts}
\usepackage{dsfont}
\usepackage{bbm}
\usepackage{colonequals}
\usepackage{amscd}
\usepackage{slashed}

\usepackage{float}
\usepackage{afterpage}

\usepackage{subcaption}

\usepackage{wrapfig}

\usepackage{suffix,mathtools,cancel,bbm}

\usepackage{multirow}

\hypersetup{
    colorlinks,%
    citecolor=blue,%
    filecolor=blue,%
    linkcolor=blue,%
    urlcolor=blue
}

\usepackage{tikz}
\usetikzlibrary{positioning}

\newcommand{\R}{{\mathbb R}}

\def\e{\epsilon}

\renewcommand{\title}[1]{\vbox{\center\LARGE{#1}}\vspace{5mm}}
\renewcommand{\author}[1]{\vbox{\center#1}\vspace{5mm}}

\makeatletter
\def\leftrightarrowsfill@{\arrowfill@\leftrarrows\Rrelbar\lrightarrows}
\newcommand{\xleftrightarrows}[2][]{\ext@arrow 3399\leftrightarrowsfill@{#1}{#2}}
\makeatother

\begin{document}

\institution{HA}{Department of Physics, Harvard University, 17 Oxford Street, Cambridge MA 02138, USA}

\institution{BHI}{Black Hole Initiative, Harvard University, Cambridge MA 02138, USA}

\institution{MI}{Dipartimento di Fisica, Universita' di Milano, via Celoria 16, 20133 Milano MI, Italy}

\institution{INFN}{INFN, Sezione di Milano, Via Celoria 16, I-20133 Milano, Italy
}

\title{Mixed 't Hooft Anomalies and the Witten Effect for AdS Black Holes}

\authors{Matthew Heydeman\worksat{\HA,\BHI} and Chiara Toldo\worksat{\HA, \MI, \INFN} \footnote[0]{mheydeman@fas.harvard.edu, chiaratoldo@fas.harvard.edu}
}

\abstract{

For a variety of BPS black holes in string theory, the supersymmetric index has provided a microscopic validation of the Bekenstein-Hawking formula. In the near-BPS limit, a gravitational path integral analysis previously revealed the semiclassical spectrum is modified, having a large extremal degeneracy (consistent with the index) and a mass gap up to a continuum of non-BPS black holes. Presently, we study examples in which these sharp features of the spectrum are altered due to the presence of anomalies in the form of $\vartheta$-angle terms in the action. These may appear generally, but we focus on near-BPS dyonic AdS$_4$ black holes in M-theory, dual to 3d $\mathcal{N}=2$ SCFTs of Class $R$ obtained by twisted compactification of $N$ wrapped M5 branes. Due to the Witten effect, the dyonic black holes receive quantum corrections to their charges, and when $\vartheta = \pi$ one may find a mixed `t Hooft anomaly between the $U(1)_R$ and $\mathbb{Z}_2$ time reversal symmetries. Using results from $\mathcal{N}=2$ JT supergravity, we find these effects result in a spectrum in which both the gap and index are reduced, and may even vanish. Surprisingly, for $\vartheta \rightarrow \pi$, neither the Bekenstein-Hawking formula nor the index correctly account for the extremal degeneracies.

}

\date{}

\maketitle

{
  \hypersetup{linkcolor=black}
  \tableofcontents
}

\pagebreak

\setcounter{page}{1}

\section{Introduction and Summary}
\label{sec:magneticAdS2}

In the passage from classical to quantum mechanics, not all properties of the classical system are in general preserved by the quantization. Energy levels and charges become quantized, and global symmetries which leave the classical Lagrangian invariant may sometimes be anomalous in the quantum theory. These properties are expected to hold for a full theory of quantum gravity, but they are not usually manifest in general relativity. Black holes in particular pose a challenge because their apparent macroscopic classical properties are in tension with their description as an ordinary quantum system. 

Following the seminal work of \cite{Strominger:1996sh}, the best understood microscopic picture of black holes involves supersymmetric (or BPS) black holes constructed in string theory via branes. However, this kind of exact microstate counting typically involves computing a supersymmetric index in a dual theory, and this index only counts ground states with a sign. Not only does the index have nothing to say about excited, non-BPS states, but it is also possible for the index to disagree with the physical degeneracies due to cancellations. Some attempts have been made to show this cancellation cannot happen in well studied examples\cite{Sen:2009vz,Mandal:2010cj,Vafa:1997gr,Benini:2015eyy}, but black holes preserving less supersymmetry may still exhibit this phenomenon.

In Lorentzian signature, such supersymmetric black holes are automatically extremal, having a macroscopic area and charges but vanishing Hawking temperature. It has recently been appreciated that as one approaches extremality, the physics depends strongly on quantum mechanical properties of the AdS$_2$ near horizon region, and these quantum effects may not be ignored when performing the semi-classical evaluation of the gravitational path integral~\cite{Iliesiu:2020qvm,Heydeman:2020hhw,Boruch:2022tno}. While these works draw lessons from the SYK model \cite{KitaevTalk,Sachdev:1992fk,Sachdev:2015efa,Maldacena:2016hyu} and JT gravity\cite{Teitelboim:1983ux,Jackiw:1984je,Almheiri:2014cka,Maldacena:2016upp,Almheiri:2016fws,Ghosh:2019rcj}, the results apply more generally for $D\geq 4$ \emph{near extremal} black holes in more general spacetimes, with or without supersymmetry\footnote{See \cite{Turiaci:2023wrh} for an overview of these developments.}. These finite temperature corrections can be seen by studying fluctuations of the full higher dimensional theory\cite{Iliesiu:2022onk,Banerjee:2023quv, Kapec:2023ruw,Rakic:2023vhv,Maulik:2024dwq,Kapec:2024zdj,Kolanowski:2024zrq}, but an intuitive and technically useful step is to reduce the higher dimensional theory to an AdS$_2$ effective gravity theory. For supersymmetric black holes, the relevant theory has been argued to be a version of JT gravity with $\mathcal{N}=2,4$ supersymmetries.

The low temperature quantum corrections (which appear due to the fluctuating boundary of JT) result in a significant departure from the Bekenstein-Hawking formula, essentially due to the one loop determinants of nearly zero modes of the metric and its superpartners. By embedding the JT gravity / Schwarzian description within a more complete bulk calculation, one can leverage the fact that the partition function of the Schwarzian theory is in fact 1-loop exact and may be computed exactly\cite{Stanford:2017thb,Mertens:2017mtv}. At sufficiently low temperatures, this quantum corrected partition function leads to a prediction for the spectrum which appears to have resolved some longstanding issues with the thermodynamic interpretation of near extremal black holes first pointed out in \cite{Preskill:1991tb}. While JT gravity is not on its own a complete description of the higher dimensional black hole, it appears as the correct effective field theory at temperatures precisely at the scale where the semiclassical analysis breaks down. 

In addition to giving a consistent picture of black hole thermodynamics at low temperatures, when supersymmetry is present, the super JT analysis is consistent with known string dualities and microstate counts at large $N$\cite{Boruch:2022tno,Iliesiu:2022kny,Boruch:2023gfn}. Beyond BPS microstate counting, the gravitational path integral and JT gravity perspectives may be used to probe other kinds of observables for supersymmetric black holes, including correlation functions of local and non-local operators and the role of wormholes\cite{Lin:2022zxd,Chen:2023lzq,Iliesiu:2021are,Boruch:2023bte} and the connection to random matrix theory and chaos \cite{Saad:2019lba,Stanford:2019vob,Turiaci:2023jfa,Johnson:2023ofr,Johnson:2022wsr,Johnson:2024tgg}. In the case of $\mathcal{N}=4$ SYM, there is some evidence that some of these gravitational effects may already be seen in the dual field theory \cite{Chang:2023zqk,Cabo-Bizet:2024gny,Chen:2024oqv}.

\subsection{Topological Terms and Anomalies}
In this article, we will study the implications of the near-BPS limit described above for a class of AdS$_4$ black holes in M-theory. In particular, we still study minimal, non-rotating, dyonic black holes\cite{Romans:1991nq,Kostelecky:1995ei,Caldarelli:1998hg} in the presence of a four-dimensional $\vartheta$ angle for the abelian gauge field under which the black hole is charged. This term is the familiar:
\begin{equation}
\label{eq:introtheta}
    S_\vartheta = \frac{i \vartheta}{8 \pi^2} \int F \wedge F \, .
\end{equation}
This term is classically $CP$ violating unless $\vartheta$ is a multiple of $\pi$, and by the $CPT$ theorem this is equivalent to a $T$ violating term. The goal of studying AdS$_4$ black holes in the presence of such a term is essentially twofold; we wish to use the solvability of the JT/Schwarzian limit to make statements about the non-BPS part of the spectrum for the dual ``topologically twisted'' 3d SCFTs\cite{Witten:1988xj,Benini:2015eyy}. Further, the presence of the $\vartheta$ angle in the bulk implies these black holes are subject to the Witten effect \cite{Witten:1979ey}, which means that quantum mechanically, their electric charges are shifted by a factor proportional to $\vartheta$. Originally understood as a phenomenon for dyons in gauge theories, we will see here (using a more careful treatment of the gravitational path integral) that the same effect applies to these black holes. This seemingly innocent shift of the charge will end up having dramatic effects on the low energy spectrum and the supersymmetric index. 

Additionally, \eqref{eq:introtheta} in 4d will end up reducing to a similar nontrivial topological term (with coefficient $\vartheta_*$ defined below) in the near horizon 2d JT supergravity, giving a precise M-theory realization of the possibility first discussed in \cite{Stanford:2017thb}. We will show in our setup that the $\vartheta_*$-angle is a tunable parameter, and a nonzero value implies both the non-BPS spectrum and the index is modified. In fact, in the limiting case of $\vartheta_* = \pi$ (mod $2\pi$), both the black hole and its field theory dual suffer from a mixed `t Hooft anomaly between $U(1)_R$ and the $\mathbb{Z}_2$ $C$ symmetry\footnote{From the point of view of JT supergravity, both $C$ and $CT$ are possible symmetries that may be anomalous\cite{Turiaci:2023jfa}. The $C$ present in the near horizon region is a remnant of $T$ from the higher dimensional point of view. While $P$ acts trivially in this version of JT, one would need to analyze this symmetry if one included matter. We thank G.J. Turiaci for comments on this point.}, even though these symmetries are both present classically and would result in an enlargement to $O(2)$. This kind of discrete mixed anomaly is not an inconsistency, but it does obstruct the existence of these symmetries in the quantum theory in the presence of a background holonomy for the electric field, similarly to \cite{Gaiotto:2017yup} for four dimensional gauge theory and \cite{Kapec:2019ecr} for JT gravity. 

The mixed anomalies in question arise for global symmetries, but there are no global symmetries in quantum gravity. Therefore, it is really the boundary (super Schwarzian) mode of JT which is the physical degree of freedom, and this theory comes with a $U(1)_R$ symmetry along with a 1-dimensional Chern-Simons term proportional to $\vartheta_*$. We will later see very concretely that the background 4d electric potential is equivalent to turning on a background gauge field for this $U(1)$, and the partition function is not invariant under the required discrete transformations with this background present.

We will elaborate on the details of the M-theory AdS$_4$/CFT$_3$ construction below, which will ultimately lead us to a class of black holes described by a version of $\mathcal{N}=2$ JT supergravity. It is important that one uses a ``mixed ensemble'' in higher dimensions where the inverse temperature and electric potential ($\beta$, $\alpha$) along with the rest of the charges are held fixed\cite{Boruch:2022tno}. In this ensemble, the JT partition function is determined by a sum over a set of gravitational saddles weighted by the one-loop determinants of a full supermultiplet of the (AdS$_2$ boundary) nearly zero modes. The result depends on a small set of discrete and continuous parameters which are determined from the higher dimensional embedding. The discrete parameters $(\hat{q},\vartheta_*)$ have a mild effect classically, but completely change the spectrum obtained from the $\mathcal{N}=2$ Schwarzian theory. The existence of these extra parameters\cite{Fu:2016vas,Stanford:2017thb,Boruch:2022tno,Turiaci:2023jfa} follows from the fact that the $\mathcal{N}=2$ Schwarzian theory possesses an abelian $U(1)_R$ symmetry; $\hat{q}$ is the unit charge of this $U(1)$ and sets the charge quantization condition, while $\vartheta$ is essentially the coefficient of a 1d Chern-Simons term which is possible in Schwarzian theories with an abelian reparametrization symmetry\cite{Kapec:2019ecr}. Both these parameters additionally have a nice interpretation in terms of $\mathcal{N}=2$ SYK models\cite{Fu:2016vas,Heydeman:2022lse,Benini:2024cpf,Heydeman:2024ohc}, where $\hat{q}$ sets the number of fermions participating in interactions, and $\vartheta_*/2\pi$ mod $1$ determines whether the total number of fermions is even or odd. 

The main conclusion from studying the spectrum of AdS$_5$ black holes using $\mathcal{N}=2$ JT gravity (dual to \emph{near} 1/16-BPS states in $\mathcal{N}=4$ SYM)\cite{Boruch:2022tno} revealed that the relevant parameters are $(\hat{q}=1,\vartheta_* =0)$. Using properties of the higher dimensional solution, this translates to a spectrum with a large extremal degeneracy (in agreement with the large $N$ superconformal index), as well as a ``mass gap'' between the BPS operators and the non-BPS operator of lowest dimension in the low temperature/extremal phase. This is qualitatively similar to the left plot of Figure~\ref{Fig0}a. However, purely from the point of view of $\mathcal{N}=2$ JT gravity, there are other more exotic possibilities for the spectrum for more general values of the parameters\cite{Stanford:2017thb}, though it was unclear if these features could be engineered in a realistic higher dimensional black hole. Here, we will answer this question for the case of the $\vartheta$ parameter, which we will show has particularly striking effects.

In order to find a simple gravity example where $\vartheta$ is non-vanishing, but there still exists a field theory quantity to compare with, we take inspiration from the possible indices that may be computed in 3d SCFT. If one chooses AdS$_4$ boundary conditions corresponding to a large Kerr-Newman black hole \cite{Kostelecky:1995ei}, the dual quantity is the large $N$ superconformal index\cite{Choi:2019zpz,Nian:2019pxj,Choi:2019dfu,Bobev:2019zmz,Benini:2019dyp,Bobev:2022wem,BenettiGenolini:2023rkq,Bobev:2024mqw}. If instead however we place the 3d SCFT on $S^1 \times \Sigma_g$, where $\Sigma_g$ is a genus $g>1$ Riemann surface with a magnetic flux for the background R-symmetry gauge field, the resulting partition function is the \emph{topologically twisted index} \cite{Benini:2015eyy,Benini:2015noa}, well known to be dual to magnetic black holes in AdS$_4$ \cite{Romans:1991nq,Caldarelli:1998hg}. Moving away from the strict BPS limit, our gravity computation can be interpreted as a prediction for the low temperature spectrum of the dual CFT$_3$ on a spatial Riemann surface $\Sigma_g$. In purely field theory terms, one might ultimately hope to compare with the models of \cite{Benini:2022bwa} for 3d $\mathcal{N}=2$ SCFTs.

The consequences of the appearance of the term \eqref{eq:introtheta} at the level of the twisted index were first studied in \cite{Choi:2020baw}. One immediate complication to studying \eqref{eq:introtheta} in a precise background is that the minimal 4d $\mathcal{N}=2$ gauged supergravity obtained by compactification of M-theory to an AdS$_4$ vacuum on a Sasaki-Einstein 7-manifold (such as $S^7$) does not actually have such a term. Therefore, the ABJM SCFT and related examples built from $N$ coincident M2 branes studied in \cite{Benini:2015eyy,Benini:2015noa} will never lead to this effect, as convincingly argued in \cite{Genolini:2021qbi}.

In fact\cite{Choi:2020baw,Genolini:2021qbi}, an appropriate 3d SCFT to consider is the low energy limit of $N$ coincident M5 branes wrapping a compact hyperbolic 3-manifold which we call $H_3$. Three dimensional superconformal field theories obtained this way are known as Class $R$ theories (denoted $T_N[H_3]$), and many of their properties may be computed using what is known as the 3d/3d correspondence \cite{Dimofte:2011ju,Dimofte:2011py,Terashima:2011qi} (see also \cite{Lee:2013ida,Cordova:2013cea,Gukov:2016gkn,Benini:2019dyp} and \cite{Dimofte:2014ija} for a review). In brief, the worldvolume theory on $N$ coincident M5 branes is the $A_{N-1}$ 6d $\mathcal{N}=(2,0)$ superconformal field theory theory, and a partial topological twist of this theory on $S^1 \times \Sigma_g \times H_3$ implies a correspondence between the twisted theory on $H_3$ and the 3d  $T_N[H_3]$ on $S^1 \times \Sigma_g$. The properties of $T_N[H_3]$ depend only on the topology of $H_3$, and in particular the index at large $N$ is then computed using the twisted theory, which is a certain $SL(N,\mathbb{C})$ complex Chern-Simons theory. At large $N$, the dual gravity system is M-theory on manifolds which are asymptotically a fibration AdS$_4$ $\times$ $H_3$ $\times$ $S^4$, where supersymmetry may be preserved by the twisting. Finally, the growth of the index is reproduced by the expected magnetic AdS$_4$ black hole solution\cite{Gang:2018hjd,Gang:2019uay}, but the phase $\vartheta$ is more subtle.

A main conclusion of our work concerns the role of $\vartheta$ for the \emph{near-BPS spectrum} which we will argue is captured by the corresponding $\mathcal{N}=2$ JT gravity. The M-theory AdS$_4$/Class $R$ duality is the sort of UV description, while the IR dynamics is the near horizon theory of the dyonic black hole, now dressed with a nontrivial $\vartheta$ topological term. The four-dimensional $\vartheta$ angle and $\vartheta_*$, the effective two-dimensional one, obey the same periodicity condition and are related in a simple way which depends only on the genus of the horizon:
\begin{align}
    \vartheta_* \equiv (1-g) \vartheta \, .
\end{align}
We will typically use $\vartheta$ to refer to the coefficient in the 4d action, and $\vartheta_*$ as the coefficient of the 2d action. The presence of this 2d term actually reduces both the gap and the index. A cartoon of the general proposal is: 
\begin{figure}[h]
	\begin{center}
	\includegraphics[width=0.80\linewidth]{"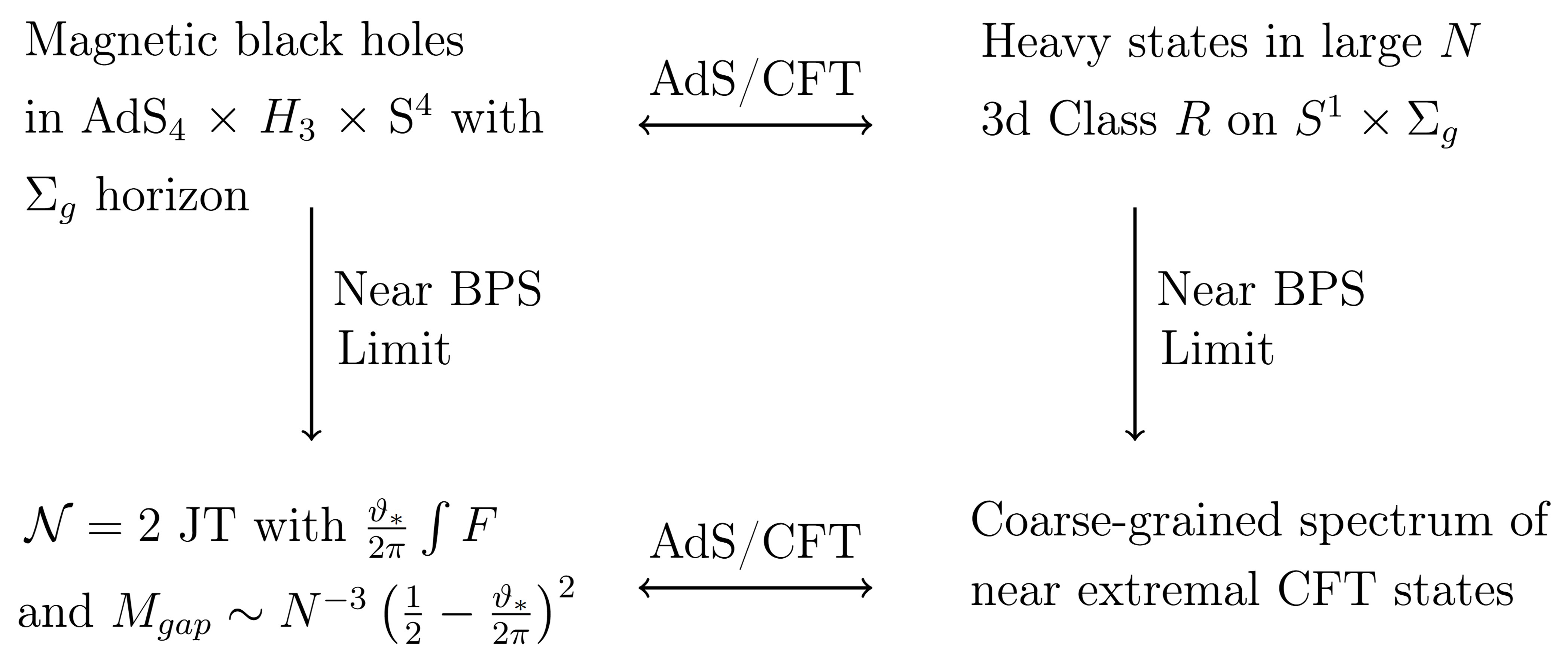"}
\end{center}
 \label{fig:twisteddiagram}
\end{figure}
\vspace{-1 cm}

Naturally, the extremal entropy, topologically twisted index, and near-BPS finite $T$ corrections obtained from this correspondence depend on the supergravity setup, and these parameters are determined by the geometry of $H_3$ $\times$ S$^4$. Crucially, the nonzero value of $\vartheta$ in the supergravity limit depends only on $N$ and a certain topological invariant of $H_3$. This hyperbolic Chern-Simons invariant takes values on the unit circle, and different choices of $H_3$ realize a ``landscape'' of possible AdS$_4$ black holes with $\vartheta$ essentially a continuous parameter.

While the black holes we consider will have vanishing electric charge classically, the charge fluctuations captured by the $\mathcal{N}=2$ Schwarzian mode will have the interpretation as small dyonic charges which are present away from extremality. Thus, even though the topological term \eqref{eq:introtheta} would seem to evaluate to $0$ on a BPS black hole with electric charge $Q=0$, it modifies the analysis of the gravitational path integral when we instead fix the potential $\alpha$ in our ensemble. This Witten effect thus must be accounted for in the path integral, which ``knows'' about fluctuations around the classical solution.

In the near-BPS limit, quantum corrections to the partition function and entropy may be efficiently computed using the JT/Schwarzian description. To see directly how JT gravity appears classically, we demonstrate the appropriate dimensional reduction for dyonic black holes in Appendix~\ref{app:dimreduction}. However, the approximate spectrum itself may only be obtained by the quantization of this theory (with disc topology via the euclidean path integral). The 1-loop exactness of this computation allows us to understand the spectrum of both the BPS and near-BPS states. The spectrum for a generic value of $\vartheta_*$ is displayed in Figure \ref{FigGen}, where both the gap and the BPS degeneracy are seen to be reduced.

For these non-BPS parts of the spectrum, it is difficult to compare our results directly with the strongly coupled field theory. However, the BPS states (being zero energy with this Hamiltonian) are also counted in the low $T$ grand canonical partition function. As we explain in Section \ref{sec:twistedindexfromJT}, a specialization of this partition function gives a gravity computation of the index, and this can be compared with the microscopic large $N$ index in field theory\cite{Choi:2020baw,Genolini:2021qbi,Gang:2019uay}. We show the gravity answer is equivalent to this microscopic result, and importantly, the index depends non-trivially on $\vartheta_*$. In fact, turning on this parameter results in an exponential reduction of the index. A suitable choice of $N$ and $H_3$ may be used to tune $\vartheta_*$ to the time reversal symmetric point $\vartheta_* = \pi$ (mod $2\pi$), and this actually causes the index to vanish completely. This can be understood already in field theory\footnote{We thank Pietro Benetti Genolini and Valentin Reys for comments on this point.}, but now we give a purely quantum gravitational explanation which can be compared with the Euclidean real solutions called ``black saddles'' \cite{BenettiGenolini:2019jdz,Toldo:2020mlu,Bobev:2020pjk} which admit a Lorentzian continuation.

 \begin{figure}[h]
	\begin{center}
	\includegraphics[scale=0.55]{"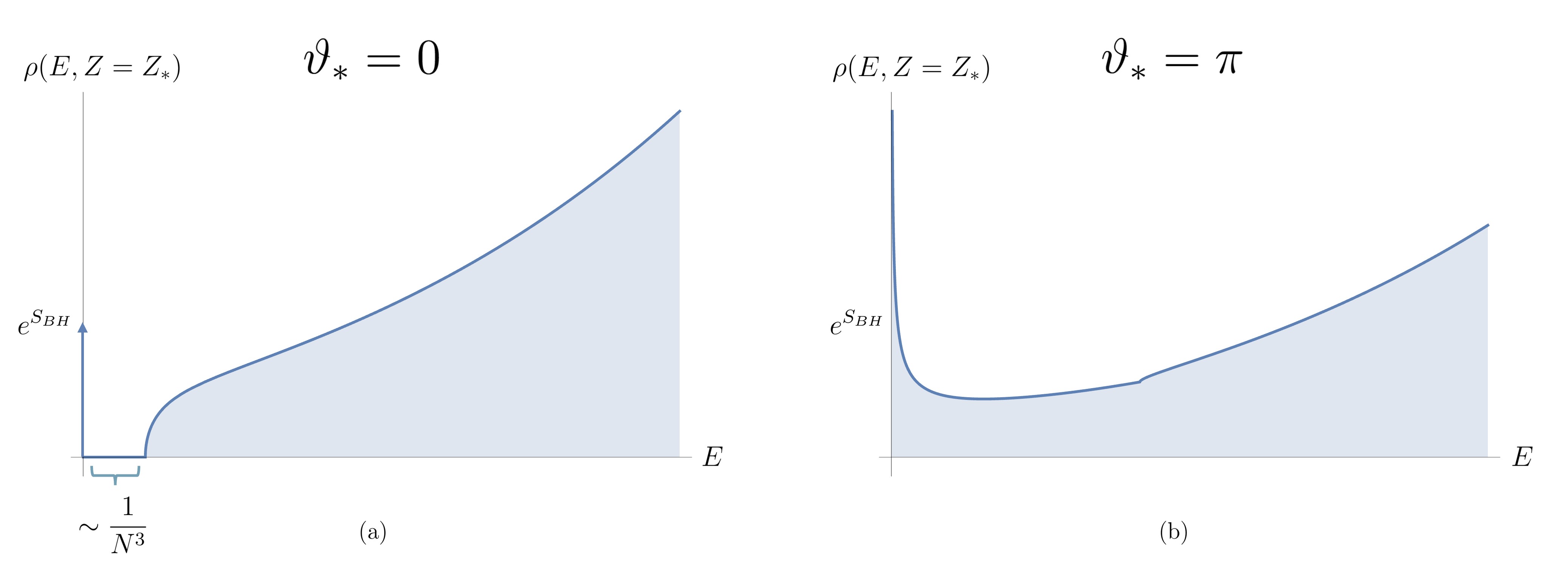"}
 \end{center}
	\caption{Graph (a) displays the density of states for a dyonic near BPS black hole in which the electric charge is equal to the BPS value ($Z=Z_*$) and the $\vartheta_*$-angle is $0$ (mod $2 \pi$). Similar to other near-BPS black holes, this displays a Dirac delta function associated to the extremal BPS solution ($S_{BH}\sim N^3$) followed by a mass gap. Graph (b) instead displays the density of states when a mixed `t Hooft anomaly is present, $\vartheta_* = \pi$ (mod $2\pi$). This non-vanishing $\vartheta_*$-angle shifts the charges of all states due to the Witten effect, which modifies the BPS bound and the gap. We have again plotted the $Z=Z_*$ charge sector, which becomes gapless, leading to an apparent divergence at small energies. } \label{Fig0}
	\end{figure}

      \begin{figure}[H]
	\begin{center}
	\includegraphics[scale=.7]{"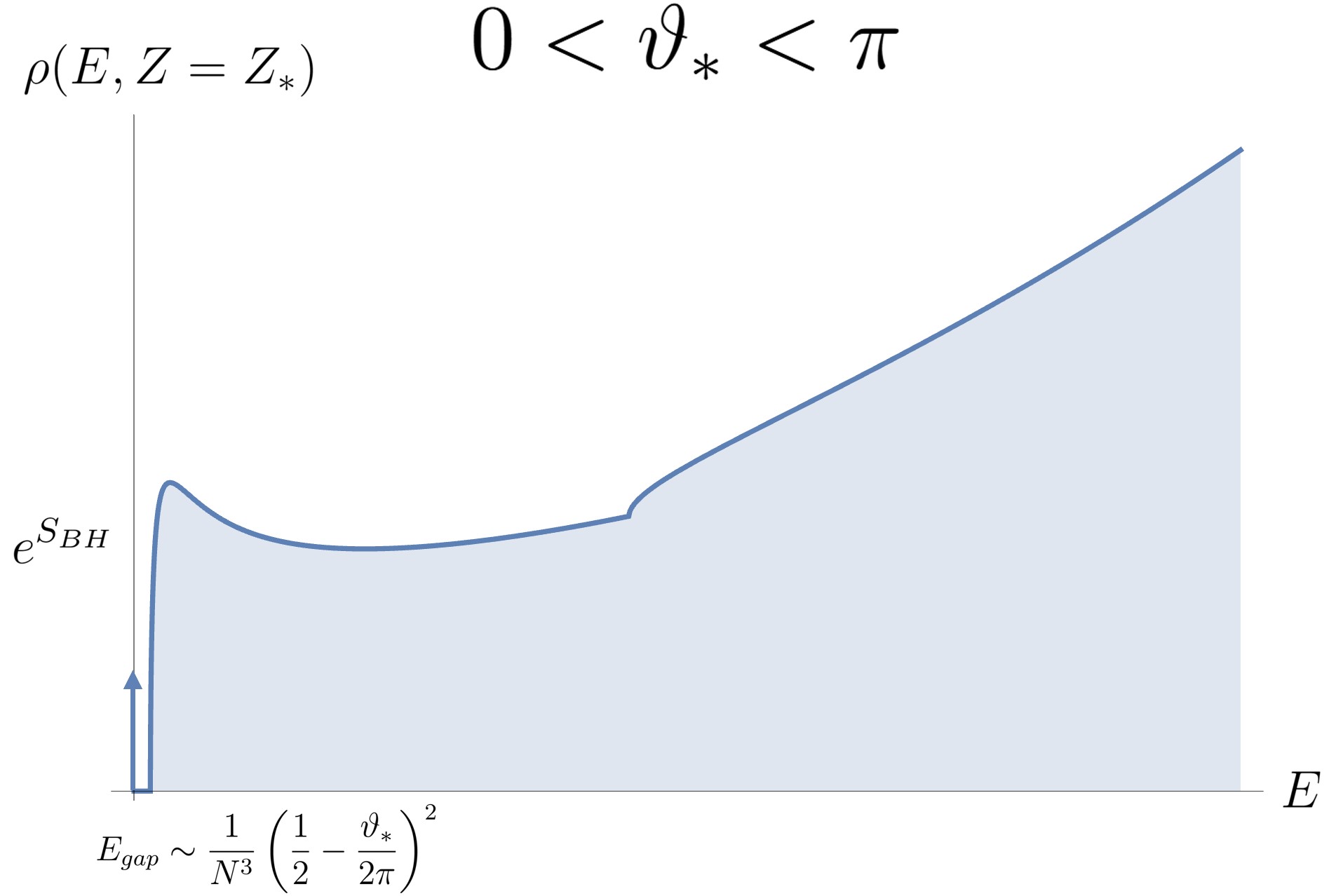"}
 \end{center}
	\caption{For generic choices of the hyperbolic 3-manifold $H_3$, the $\vartheta_*$ angle is essentially a continuous parameter that is $2\pi$ periodic. This figure shows a representative density of states for a dyonic near BPS black hole in which the electric charge is equal to the BPS value ($Z=Z_*$) and the $\vartheta_*$-angle falls in the range $0< \vartheta_* < \pi$. It displays an interpolating behavior between the two limiting cases of Figure \ref{Fig0}. The gap closes as a polynomial in $\vartheta_*$, and the BPS degeneracy (represented by the delta function) is reduced.} \label{FigGen}
	\end{figure}

The fact that the index may vanish seems to be a paradox from the gravity point of view. The black hole has a large area, but the analytically continued complex saddles of \cite{Choi:2020baw,Genolini:2021qbi} come with an additional phase which can cause this partial or complete cancellation. Thus, the fate of the field theory partition function (rather than the index) as $\beta \rightarrow \infty$ is obscured in this approach. Using the gravity computation of the partition function, we instead find that a set of long (non-BPS) multiplets maintains an exponentially large degeneracy as the energy $E\rightarrow 0$ (see the second graph of Fig. \ref{Fig0}b). This results in a large discrepancy between the Bekenstein-Hawking formula (due to the 1-loop corrections near extremality), the topologically twisted index (due to its vanishing), and the exact answer from gravity.

 Our work shows at large $N$ that this cancellation is not due to Bose-Fermi degeneracy of the ground states, but rather because a version of the Witten effect shifts the BPS charges outside of the BPS range allowed by $\mathcal{N}=2$ JT gravity. We also may understand the spectrum of long multiplets in the theory, and in fact the gap closes due to a single set of multiplets which move towards but do not saturate the BPS bound. The same behavior has been found in the context of $\mathcal{N}=2$ SYK models with an odd number of fermions \cite{Stanford:2017thb}, and it is interesting to find it realized here in the context of black holes with a particular eleven-dimensional origin. To sum up, with reference to the Figure \ref{Fig0} and \ref{FigGen}:
\begin{itemize}
\item We find that the BPS and near-BPS spectrum of dyonic AdS$_4$ black holes (discussed classically in Section \ref{Sec:dyonicsolutions})  depends sensitively on the value of the 2d $\vartheta_*$-angle, and this parameter in turn depends on the compactification from 11d.  For $\vartheta_* =0$ (relevant for ABJM and Class $R$ theories with $cs(H_3)=0$), we find the spectrum is similar to the one for AdS$_5$ black holes\cite{Boruch:2022tno}. When the electric charge is set to the BPS value, this spectrum has a BPS delta function (reproducing the extremal Hawking result) and a mass gap followed by a continuous density of states.
\item For Class $R$ theories dual to (twisted) AdS$_4 \times H_3 \times S^4$ vacuua, a generic $H_3$ realizes essentially a continuous value of $\vartheta_*$. This parameter enters into the $\mathcal{N}=2$ JT supergravity path integral much like  the extremal entropy $S_*$, and the details of the spectrum are discussed in Section \ref{sec:nearBPSDOS}. The anomaly modifies the sum over $U(1)_R$ bundles in the grand canonical JT calculation; it also results in a Witten effect which shifts $U(1)_R$ charges of all states by $\vartheta_*$. The density of states extracted from the JT path integral now has a mass gap that depends on the value of $\vartheta_*$, with the minimal gap scaling as $E_{gap} \sim \frac{1}{N^3}\left(\frac12 -\frac{\vartheta_*}{2\pi} \right)^2$. The BPS degeneracy manifests as a delta function at the origin, though the number of states is reduced by a factor $\cos \left(\frac{\vartheta_*}{2} \right)$. This degeneracy is in agreement with the microscopic index at large $N$ in field theory \cite{Choi:2020baw,Genolini:2021qbi}.
\item At the point $\vartheta_* = \pi$, there is a classical $CT$ symmetry, but this symmetry suffers a mixed anomaly with the global $U(1)_R$ transformations of the boundary, as can be seen explicitly from our partition function. This behavior in principle can be realized by certain choices of $H_3$, and our general formula for the spectrum reveals the index actually vanishes at this anomalous point. This vanishing is not due to supersymmetric cancellations, but rather because the Witten effect shifts all charges outside the allowed BPS bound. In the sector of charge fixed to the BPS value, the states are gapless and there is a large (divergent) number of semiclassical non-BPS states near $E=0$, as explained in Section \ref{ssec:anomalousdegeneracy}. Therefore, at the anomalous point we must be careful in interpreting the one-loop gravity answer and draw a distinction between degenerate BPS ground states and course-grained near ground states. This suggests that both the Hawking formula and microscopic index do not correctly account for the degeneracies due to the mixed anomaly. 
\end{itemize}

\section{Magnetic and Dyonic Solutions}
\label{Sec:dyonicsolutions}
As we have explained in the introduction, our goal is to study the quantum spectrum of dyonic AdS$_4$ black holes in the presence of a four-dimensional $\vartheta$-angle. Based on the arguments reviewed in \cite{Genolini:2021qbi}, one way to achieve the AdS$_4$ vacuum is to start with M-theory in the presence of $N$ M5-branes wrapping a general compact hyperbolic three-manifold $H_3$, leading to 3d theories of Class $R$\cite{Choi:2020baw}. 

Starting from eleven dimensional supergravity, one proceeds in two steps. First, one reduces to AdS$_7 \times S^4$ via a suitable ansatz, obtaining maximal $SO(5)$ gauged supergravity in 7d. One then is interested in seven dimensional solutions that contain a $H^3$ factor: imposing a further reduction ansatz one lands on minimal $\mathcal{N}=2$ supergravity with a topological term of the form \eqref{eq:introtheta}, which comes directly from the reduction of the topological term present in the 7d theory. We refer the reader to  \cite{Genolini:2021qbi} for the details of the truncation.  From the bulk point of view\cite{Pernici:1984xx,Pernici:1984nw}, the 7d internal space is an $S^4$ bundle over $H_3$. The four-dimensional minimal supergravity with a $\vartheta$ term for the gauge field reads
\begin{equation}
\label{eq:einsteinmaxwelltheta}
    S_{4d, \vartheta} = -\frac{1}{16 \pi G} \int d^4x \sqrt{g} \left(R + 6 - F^2 \right) + \frac{i \vartheta}{8 \pi^2} \int F \wedge F \, .
\end{equation}
The $\vartheta$-term and Newton constant are given in terms of the 11d background data, at leading order in $N$, by topological invariants of the compact hyperbolic space $H_3$,
\begin{equation}
\label{eq:vartheta11d}
\frac{1}{G} = \frac{2 \vol(H_3)}{3 \pi^2} N^3\,, \qquad  \qquad  \vartheta = \frac{2 cs(H_3)}{3}N^3 \, ,
\end{equation}
where $N$ is the number of M5-branes, and $cs(H_3)$ is the Chern–Simons invariant of the
hyperbolic $H_3$:
\begin{align}
\label{eq:CSH3}
    cs\left(H_3\right)=\frac{1}{8 \pi} \int_{H_3} \textrm{Tr}_3\left(\omega \wedge \mathrm{~d} \omega+\frac{2}{3} \omega \wedge \omega \wedge \omega\right) \, ,
\end{align}
where $\omega$ is the spin connection and the trace is in the defining representation. As discussed in \cite{Choi:2020baw,Genolini:2021qbi}, this invariant is only defined mod $2\pi$. Because of this, for the 4d $\exp(\frac{i \vartheta}{8 \pi^2} \int F \wedge F)$ term to be well defined in the path integral, we actually require that $\vartheta$ be periodic with period $2\pi$. This is clearly not satisfied for the above expression for generic $N$. The resolution is that the exact answer for $\vartheta$ receives subleading corrections in $N$; the full form of these are unknown, but the first correction was determined in part in \cite{Genolini:2021qbi}\footnote{Importantly, the quantization condition for the M-theory antisymmetric tensor receives gravitational contributions\cite{Witten:1996md}.} to be $\vartheta = cs(H_3) \frac{2N^3-N}{3} + \mathcal{O}(N)$. We ignore this subtlety for now, but the exact mod $2\pi$ quantization of $\vartheta$ will play an important role later.

The $\vartheta$-term in the action \eqref{eq:einsteinmaxwelltheta} is topological, hence it does not show up in the equations of motion. The solutions to the equations of motion are then the ones found in the usual setups of $\mathcal{N} =2$ minimal gauged supergravity with $\vartheta =0$. We describe here the general black hole solutions of minimal gauged supergravity which allow for supersymmetric limits \cite{Romans:1991nq, Caldarelli:1998hg}. We start from metric and gauge field
\begin{equation}
\label{eq:Minimal_StaticDyon}
     ds^2 = - V(r) \, d t^2 + \frac{d r^2}{V(r)} + r^2 \left( d\theta^2 + \sinh^2\theta \, d \phi^2 \right) \, ,
    \end{equation}
\begin{equation}
    \cA = \frac{Q}{r} \, d t + P \cosh\theta \, d\phi + \gamma_t \, d t \, ,
\end{equation}
where $\gamma_t$ is a constant and 
\begin{equation}
\label{eq:Minimal_StaticDyon_Vr_v1}
    V(r) = r^2 -1 - \frac{2M}{r} + \frac{Q^2 + P^2}{r^2} \, .
\end{equation}
We use the coordinates $\theta$, $\phi$ to write a local form of the constant curvature metric on the genus-$g$ Riemann surface $\Sigma_g$, normalized so that $\text{vol}(\Sigma_g) = 4\pi (g-1)$ for $g>1$.

This solution describes a family of asymptotically locally AdS static dyonic black holes with $\Sigma_g$ horizon located at $r_h$, the largest positive root of $V(r)$. The null generator of the horizon is the Killing vector $\xi=\partial_t$. The entropy of the horizon is given by the Bekenstein--Hawking formula
\begin{equation}
\label{eq:Minimal_StaticDyon_Entropy}
    S = \frac{r_h^2}{4G} \text{vol}(\Sigma_g) \, .
\end{equation}
The electrostatic potential of the solution is
\begin{equation}
\label{eq:Minimal_Phi}
    \Phi \equiv \iota_\xi \cA|_{r=r_h} - \iota_\xi \cA|_{r\to \infty} = \frac{Q}{r_h} \, .
\end{equation}
We perform a Wick rotation by defining $t = -i \tau$. Regularity of the metric allows us to compute the temperature of the horizon as
\begin{equation}
\label{eq:Minimal_StaticDyon_Temperature}
    \beta = \frac{4\pi}{V'(r_h)} \, ,
\end{equation}
and regularity of the gauge field at the horizon requires $\iota_\xi \cA|_{r=r_h} = 0$, so that $\gamma_t = -Q/r_h = -\Phi$.
The resulting solution is a Euclidean metric on $\R^2\times \Sigma_g$, and an imaginary gauge field which can be made real defining $Q = i \tilde{Q}$.

The holographic charges of the solutions, computed for instance as in \cite{BenettiGenolini:2023ucp}, are
\begin{equation}
\label{eq:Minimal_StaticDyon_Charges}
\begin{aligned}
    q = \frac{Q}{4\pi G} \text{vol}(\Sigma_g) \, , \qquad  p = - \frac{P}{4\pi G} \text{vol}(\Sigma_g) \, , \qquad
    E = \frac{M}{4\pi G} \text{vol}(\Sigma_g) \, , 
\end{aligned}
\end{equation}
and the holographically renormalized Euclidean action is obtained by plugging the solution into the action \eqref{eq:einsteinmaxwelltheta} and renormalizing it via standard  techniques 
\begin{eqnarray} \label{eq:Minimal_StaticDyon_QuantumStatistical_v1theta}
    I &= & - \frac{\beta}{8\pi  G} \left( r_h^3 - M + \frac{Q^2-P^2}{r_h} \right) \text{vol}(\Sigma_g) \, + \frac{\ \vartheta \beta \, \Phi \, P }{4\pi^2}  \text{vol}(\Sigma_g) \nonumber \\
    & = &  -S + \beta (E - \Phi q) + \frac{\ \vartheta \beta \, \Phi \, P }{4\pi^2}  \text{vol}(\Sigma_g) \, .
\end{eqnarray}

\subsection{Effect of the $\vartheta$-angle}

We now comment on the interpretation of the topological term,
\begin{equation}
    S_\vartheta =  \frac{i \vartheta}{8 \pi^2} \int F \wedge F \, .
\end{equation}
We are instructed to evaluate this term on a manifold that is asymptotically AdS$_4$ with certain thermal boundary conditions. Note that if we were instead considering a compact spin manifold without boundary, then this term is always an integer, $i S_\vartheta|_{\textrm{compact}} \in \mathbb{Z}$. However, our AdS space has a boundary $S^1 \times \Sigma_g$, and therefore the topological term becomes a pure boundary term:
\begin{equation}
\label{eq:chernsimonsaction}
    S_\vartheta =  -\frac{i \vartheta}{8 \pi^2} \int_{S^1 \times \Sigma_g} A \wedge dA \, .
\end{equation}
It is argued in \cite{Genolini:2021qbi} that this finite counterterm should be added to the action of the 3d Class $R$ SCFT. From this field theory path integral perspective, we are computing the partition function in the presence of a background $A$ flux and holonomy, where $A$ is a background connection for the $U(1)_R$ symmetry. If this is done via saddle point (either because we are evaluating the index via localization, or a large $N$ expansion of the path integral), then this boundary term attaches an additional phase to different saddles which depend only on the boundary conditions for $A$. In our ensemble, we will ultimately fix the holonomy of $A$.

Related to the discussion above, we can consider the action of the transformation $\vartheta \rightarrow \vartheta + 2\pi$. This can be understood as the $T$ transformation of the $SL(2,\mathbb{Z})$ electric-magnetic duality group enjoyed by Eq. \eqref{eq:einsteinmaxwelltheta}. These bulk duality transformations have a natural interpretation in 3d CFT\cite{Witten:1995gf}. While in the bulk, we have $S$-duality, from the boundary point of view these are not duality transformations, but rather an operation which modifies the CFT by changing the background fields. Under $\vartheta \rightarrow \vartheta + 2\pi$, the field theory partition function changes as
\begin{align}
\label{eq:Z3dcounterterm}
    Z_{3d}[A,\, S^1\times\Sigma_g] \rightarrow Z_{3d}[A,\, S^1\times\Sigma_g]e^{-i S_{CS}(A)} \, .
\end{align}
Namely, we shift the partition function by a level $1$ Chern-Simons term, where $S_{CS}(A)$ takes the standard form as in \eqref{eq:chernsimonsaction} with unit coefficient. Considering the 3d SCFT with the additional counterterm has the effect of fixing our boundary conditions and removing this ambiguity. For generic $\vartheta$, the fractional part of this counterterm (mod $2\pi$) is physical and may strongly modify the dynamics\cite{Closset:2012vp}.

In this article, we are largely concerned with the implications of the $\vartheta$ angle on the spectrum of near-BPS black holes. Classically, one must use an imaginary gauge field or more generally a complex metric to describe these minimal dyonic solutions as in \eqref{eq:Minimal_StaticDyon_Charges}. The effect of this is embodied in \eqref{eq:Minimal_StaticDyon_QuantumStatistical_v1theta}, which is imaginary in Euclidean signature. We will understand how this additional topological term modifies the quantum theory of fluctuations in the near horizon region by performing the semiclassical analysis around the BPS limit.

\subsection{Expansion around the BPS limit}

In mimimal $\mathcal{N}=2$ gauged supergravity the bosonic action \eqref{eq:einsteinmaxwelltheta} is augmented by additional gravitino terms of the general form found in \cite{Freedman:1976aw,Fradkin:1976xz,Romans:1991nq}. At the level of the classical solution, all the fermions are set to zero. Nevertheless, supersymmetry of a solution is diagnosed by the existence of a supercovariantly constant Killing spinor, which is a spinor annihilated by the local supersymmetry variation of the gravitino:
\begin{align} \label{Killing-spinor}
    \left (\nabla_\mu - i A_\mu +\frac12 \Gamma_\mu +\frac{i}{4}F_{\nu \rho}\Gamma^{\nu \rho}\Gamma_\mu \right ) \epsilon = 0 \, ,
\end{align}
where
\begin{equation}
\nabla_{\mu} \epsilon = \partial_{\mu} \epsilon - \frac14 w_{\mu}^{ab} \gamma_{ab} \epsilon \, ,
\end{equation}
One important observation from \cite{Heydeman:2020hhw,Boruch:2022tno}  is that gravitinos contribute additional fluctuation determinants which can not be ignored in the low temperature limit\footnote{In studying quantum corrections more generally, one might complain that this logic should apply to all fluctuating fields of the 11d metric more generally, and not just the components that appear in our truncation. One can argue \cite{Iliesiu:2020qvm,Iliesiu:2022onk}  that near extremality, these additional fluctuations contribute to the $\log(S)$ corrections to the entropy. Since we want to focus on the $\log(T)$ corrections only, we will work at leading order in the extremal entropy $S_0$ with the understanding that additional corrections may be included in a systematic way, see for instance \cite{Sen:2012kpz} and related work. }. To determine the form of these corrections, we first study the strict BPS limit, which determines which supersymmetry algebra is broken at finite temperature (similar to the soft breaking of the AdS$_2$ $SL(2,\mathbb{R})$ symmetry at finite temperature\cite{Maldacena:2016hyu}). The set of broken reparametrization modes essentially labels which version of the Schwarzian/JT model to use, including the gravitinos. An example of this is explained systematically in \cite{Heydeman:2020hhw}, and we follow a similar route but omit some details.

Thus the fermions (while classically trivial) actually play an essential role in our later results. Indeed, quantum corrections to the black hole thermodynamics near extremality are qualitatively very different depending on whether we just consider a generic bosonic cosmological Einstein-Maxwell bosonic theory such \eqref{eq:einsteinmaxwelltheta} or if we consider the $\mathcal{N} =2 $ theory that contains gravitinos as well.  We focus on this second option.

The black holes \eqref{eq:Minimal_StaticDyon} are supersymmetric provided \cite{Caldarelli:1998hg}
\begin{equation}
\label{eq:Minimal_StaticDyon_SUSY}
   M = 0 \, , \qquad P^2 = \frac{1}{4} \, ,
\end{equation}
with electric charge unconstrained. The warp factor in this case reads
\begin{equation}
\label{eq:twistedharmonicfunction}
U(r) = r^2 -1 + \frac{1}{4r^2} + \frac{Q^2}{r^2} = \left(r-\frac{1}{2r} \right)^2 + \frac{Q^2}{r^2} \, .
\end{equation}
The solution has a (double) horizon if the electric charge vanishes, and it is a naked singularity as soon as $Q$ is real and nonzero.  The Killing spinor for this solution, in the conventions of \cite{BenettiGenolini:2023ucp} is 
\begin{equation}\label{KSfullsol}
    \epsilon = \e^{\frac{i\, t}{2} \gamma_t} \left(\sqrt{f_1(r) - f_2(r)} - i \Gamma_0 \sqrt{f_1(r) +f_2(r)} \right) \left( \frac{1-\Gamma_1}{2}\right)\left( \frac{1-i \Gamma_2 \Gamma_3}{2}\right) \epsilon_0 \, ,
\end{equation}
with
\begin{equation}
f_1(r) = \frac{\sqrt{\left(r^2- \frac{1}{2}\right)^2+Q^2}}{r}\,, \quad \qquad f_2(r) = \frac{r^2 - \frac12}{  r} \,,
\end{equation}
and $\epsilon_0$ a constant spinor \cite{Romans:1991nq,Caldarelli:1998hg}. Notice that, due to the double projection in \eqref{KSfullsol}, the solution is 1/4 BPS.
We have adopted the same conventions as those in \cite{BenettiGenolini:2023ucp} for Gamma matrices and supersymmetry conditions.

The extremal solution, which has $Q=0$, has $AdS_2 \times \Sigma_g$ near horizon geometry:
\begin{equation}
ds^2 = - 2\rho^2 dt^2 +\frac{d\rho^2}{2 \rho^2} + \frac12 (d\theta^2 + \sinh^2 d\phi^2)\,, \qquad F = P \sinh \theta \,.
\end{equation}
Such near horizon geometry is 1/2 BPS, in the sense that there is an enlargement by the additional conformal supersymmetry. To see this one can see that the Killing spinor equations are satisfied by the following pair of spinors
\begin{eqnarray}
\epsilon_1 & = &  \sqrt{\rho} \left( \frac{1-\Gamma_1}{2}\right)\left( \frac{1-i \Gamma_2 \Gamma_3}{2}\right) \epsilon_0  \, ,\\
\epsilon_2 & = & \left(-\frac{1}{2 \sqrt{\rho}} + 2 \,t \sqrt{\rho} \, \Gamma_0 \right) \left( \frac{1+\Gamma_1}{2}\right)\left( \frac{1-i \Gamma_2 \Gamma_3}{2}\right) \epsilon_0 \, ,
\end{eqnarray}
where $\epsilon_0$ is again a constant spinor, unconstrained. Each one of $\epsilon_{1,2}$ has two projectors that reduce the supersymmetries to 1/4, so each one of them has two independent components, for a total of 4 supersymmetries preserved (hence the $AdS_2 \times \Sigma_g$ solution is 1/2 BPS). The superalgebra of the horizon is $SU(1,1|1)$, as shown in \cite{Hristov:2013spa}: SU(1,1) being the isometries of AdS$_2$ and U(1) is the R-symmetry. Notice that due to the quotient to produce a higher genus Riemann surface, the isometry group SO(2,1) of 2d hyperbolic space is generically broken.  

We will focus our attention on specific configurations called ``black saddles'', first treated in \cite{Bobev:2020pjk}. They are found by performing the Wick rotation $t \rightarrow i \tau$, and also by imposing a purely imaginary value of the electric charge $Q$, such that $Q = i \tilde{Q}$ with $\tilde{Q} \in \mathbb{R}$. The metric and field strength are real and read
\begin{equation} \label{sol_rot_eucl}
    ds^2 = V(r) d\tau^2 + \frac{dr^2}{V(r)} +r^2 (d\theta^2 + \sinh^2\theta d\phi^2) \, , \qquad F = \frac{\tilde{Q}}{r^2} d\tau \wedge dr + \frac12 \sinh \theta d\theta \wedge d\phi \, ,
\end{equation}
with
\begin{equation}
V(r) = r^2 -1 + \frac{1}{4r^2} - \frac{\tilde{Q}^2}{r^2} = \left(r-\frac{1}{2r} \right)^2 - \frac{\tilde{Q}^2}{r^2} \, .
\end{equation}

These configurations can be regarded as well as the $p=0$, $g>1$ version of the solutions first found in \cite{Toldo:2017qsh,BenettiGenolini:2019jdz}. Notice that the metric is regular and caps off smoothly\footnote{These configurations are also named ``Bolt'', in the language introduced for instance in \cite{Martelli:2012sz}, because the Killing vector $\partial_t$ has a two dimensional fixed point (in contrast to the ``NUT'' solution for which such fixed point is zero dimensional).} at $r= r_0 \equiv \sqrt{\frac12 +|\tilde{Q}|}$ provided that $\tau \sim \tau + \frac{4 \pi}{|V'(r_0)|}$. If $\tilde{Q}=0$, the metric develops an infinite throat and can be Wick-rotated into a regular black hole solution, with horizon at $r_0 = \frac{1}{\sqrt2}$.

In the supersymmetric limit, starting from \eqref{eq:Minimal_StaticDyon_QuantumStatistical_v1theta} we have 
\begin{equation}
I|_{SUSY, \vartheta} = -S +\beta (\Phi-\Phi_*) q -  \frac{\beta \vartheta}{2\pi} (\Phi- \Phi_*) = I|_{SUSY} - \frac{\beta \vartheta}{2\pi} (\Phi- \Phi_*) (g-1) \,,
\end{equation}
where we took into account that in the susy limit $E=0$ and $P= 1/2$, and $\Phi_* =0$ since the extremal BPS solution has zero electric charge. For simplicity we choose the plus sign in the value of the magnetic charge, the negative branch can be treated in the analogous way. The expression for $I|_{SUSY}$ is the one obtained in \cite{BenettiGenolini:2023ucp} considering the $\vartheta =0$ setup. Notice that this formula, if computed explicitly on the Euclidean solution \eqref{sol_rot_eucl} with $\tilde{Q} \neq 0$, leads to 
\begin{equation}\label{FE_SUSY_twisted}
I|_{SUSY, \vartheta} = - \left( \frac{\pi}{2 G} + \frac{i \vartheta}{2} \text{sgn}(\tilde{Q}) \right) (g-1) \, ,
\end{equation}
which is indeed the result found in \cite{Genolini:2021qbi}. This ultimately arises because the extremal BPS limit ($\tilde{Q} \rightarrow 0$) is not smooth, since the limits $\tilde{Q} \rightarrow 0^+$ and  $\tilde{Q} \rightarrow 0^-$ do not coincide. Indeed the extremal (purely magnetic) configuration can be reached from two different Bolt solutions, in the limit in which the Bolt recedes to infinite distance \cite{BenettiGenolini:2019jdz,Toldo:2020mlu}.

Starting from here,
we can then expand the expression for the free energy 
in series for small $T$ and defining the conventional variable $\varphi \equiv  \Phi $, the free energy is then 
\begin{align}
    I(\beta,P=\frac12, \varphi) = \left( -\frac{1}{8 G} + \frac{\beta \vartheta \varphi}{8 \pi^2 } - \frac{1}{16\sqrt{2} G \pi} \frac{2\pi^2}{\beta}(1+\frac{\beta^2 \varphi^2}{\pi^2 })\right )\vol(\Sigma_g) \, .
\end{align}
Defining $\alpha$ as
\begin{align}
   \alpha \equiv \frac{\beta \varphi }{2\pi i} \, ,
\end{align}
we have
\begin{align}
    I = \left( -\frac{1}{8 G} + \frac{i \alpha \vartheta }{4 \pi} - \frac{1}{16\sqrt{2} G \pi} \frac{2\pi^2}{\beta}(1-4\alpha^2)\right )\vol(\Sigma_g)\,.
\end{align}
We now set $\vol(\Sigma_g) = 4\pi (g-1) $, giving
\begin{align}
    I = \left( -\frac{\pi (g-1)}{2 G} + i  (g-1) \alpha \vartheta  - \frac{ (g-1)}{4\sqrt{2} G } \frac{2\pi^2}{\beta}(1-4\alpha^2)\right )\, .
\end{align}
The first two terms of this expression represent the action of the extremal solution in this ensemble, and the third term gives the leading correction in the temperature. Following the arguments of \cite{Boruch:2022tno}, it can also be regarded as the classical action of $\mathcal{N}=2$ JT gravity with a gauge field and our chosen boundary conditions. The action can be recast in a convenient form by making the following definitions:
\begin{align}
\label{eq:nearBPSDefs}
    S_* \equiv \frac{\pi (g-1)}{2 G} \, , \qquad \vartheta_* \equiv (1-g)\vartheta \, , \qquad M^{-1}_{gap} \equiv \frac{(g-1)}{4 \sqrt{2} G} \, ,
\end{align}
we have
 \begin{equation} \label{FE_exp_mix}
   I (\beta, P =\frac12, \alpha) =  -S_* - i \alpha \vartheta_*    - \frac{ 2\pi^2}{\beta M_{gap}} \left( 1-4 \alpha^2 \right) \, .
\end{equation}
Once again, as a cross check, we note that if we use the susy value $\alpha = \frac12$ and take minus the action, we get using \eqref{eq:vartheta11d},
\begin{align}
    -I_{susy}= \frac{(g-1)N^3}{3\pi}\left (\vol(H_3) - i \pi cs(H_3) \right ) \, .
\end{align}
which matches the value reported in \cite{Genolini:2021qbi}. As explained in the introduction, We will use $\vartheta_*$ throughout the rest of the draft to denote the effective 2d $\vartheta$ angle, and reserve $\vartheta$ for the 4d one.

\section{The near BPS density of states}
\label{sec:nearBPSDOS}
In the previous section, we have argued that the extremal magnetic hyperbolic black hole has $M=0$, $Q=0$, $P=\frac12$ and an AdS$_2 \times \Sigma_g$ near horizon geometry. This solution possesses two Killing spinors which are constant on $\Sigma_g$, and the resulting symmetry algebra of the solution is $SU(1,1|1)$, where the $U(1)_R$ part of this is identified with 4d electric gauge transformations. This is in contrast with the 4d $\mathcal{N}=2$ Reissner-Nordstrom solution with $PSU(1,1|2)$ symmetry\cite{Heydeman:2020hhw} and the AdS$_4$/AdS$_5$ Kerr-Newman solution with $SU(1,1|1)$ symmetry \cite{Boruch:2022tno}. In these cases, the Killing spinors are nontrivial function on the angular directions parameterizing the horizon, and thus the respective $G_R \equiv SU(2)_R$ or $U(1)_R$ symmetries involve a certain linear combination of electric gauge transformations and angular isometries of the horizon. When we move away from the extremal limit, the supergroup containing $SL(2,\mathbb{R})\times G_R$ is weakly broken, and our choice of how to move away from the extremal limit is parameterized by our choice of ensemble. In all cases the form of the internal R-symmetry in part determines our choice of mixed ensemble, where we have potentials $\beta$ but also $\alpha$, which is the holonomy at the AdS$_2$ boundary for the $G_R$ symmetry. The remaining charges are held fixed. In situations where we specialize to an index with an exact field/string theory computation, it has been argued that there exists a family of (complex euclidean) saddles with different winding numbers $m_i$ for the Cartan sublagebra of the internal symmetry group \cite{Aharony:2021zkr,BenettiGenolini:2023rkq,Iliesiu:2022kny}, and these correspond to the sum over saddles found in the effective AdS$_2$ JT gravity computation.

In the present example of dyonic AdS$_4$ black holes in the classical theory, we have found the near-BPS expansion of the action in the mixed ensemble where $\alpha$ is the holonomy of the 4d electric gauge field at the boundary. Similar to better studied cases discussed above, we propose a corresponding sum over Euclidean saddles with different winding numbers $m$ for this gauge field. This sum, combined with a generalization of \eqref{FE_exp_mix} for the shifted saddles, allows us to conjecture a form of the partition function near the BPS limit similar to what was argued for in the Kerr-Newman case in \cite{Aharony:2021zkr,Boruch:2022tno}:
\begin{align}
\label{eq:conjectureZtwisted}
    Z \sim \sum_{m \in \mathbb{Z}} Z_{\textrm{one-loop}}e^{-I_{ME}(\beta, \alpha+m)} \, .
\end{align}
Given this proposal, combined with the fact that the effective near-AdS$_2$ theory is again $\mathcal{N}=2$ JT gravity based on the breaking of $SU(1,1|1)$, which we argue for in Appendix \ref{app:dimreduction}, we will use this JT/Schwarzian description to compute $Z_{\textrm{one-loop}}$ and interpret the result.

We are not aware of any microscopic analysis for the specific example of AdS$_4 \times H_3 \times S^4$ twisted black holes that discusses these subleading saddles from a bulk or boundary point of view at the level of \cite{Aharony:2021zkr}. However, note that there has been some work \cite{Benini:2022bwa} on the field theory description of these twisted black holes away from the supersymmetric limit, at least for theories related to ABJM. For 3d $\mathcal{N}=2$ Class R theories at large $N$, our main point of comparison is the computation of the topologically twisted index \cite{Gang:2019uay,Choi:2020baw,Genolini:2021qbi}, which involves a sum over the pair of complex saddles \eqref{FE_SUSY_twisted}. Readers skeptical of our proposal for the non-protected \eqref{eq:conjectureZtwisted} using the JT gravity description and the sum over saddles may wish to look at Section \ref{sec:twistedindexfromJT}, where we show that the partition function \eqref{ZJT3} we propose below correctly reproduces the large $N$ index of \cite{Choi:2020baw,Genolini:2021qbi} by taking the supersymmetric limit of the JT answer ($\alpha = \frac12$).

For convenience, we recall again the disc partition function of $\mathcal{N}=2$ JT gravity, which is a specific theory whose partition function function takes the form of \eqref{eq:conjectureZtwisted}\cite{Stanford:2017thb,Mertens:2017mtv}:
\begin{equation}\label{FE_JT_with_theta_s6}
    Z_{\mathcal{N}=2 \textrm{ JT}}(\beta, \alpha ; r, \vartheta_*) = \sum_{m \in \frac{1}{r}\cdot \mathbb{Z}} e^{i r \vartheta_* m} \frac{2 \cos(\pi (\alpha + m))}{\pi (1-4(\alpha+m)^2)}e^{S_* + \frac{2\pi^2}{\beta M_{gap}}(1-4(\alpha +m)^2)} \, .
\end{equation}
This expression depends only on information in the AdS$_2$ region of the geometry, and depends on the parameters $S_*$, $M_{gap}$, $\vartheta_*$ as well as the discrete parameter $r$. We will see that $r=1$ is consistent with the known answer for the field theory index in \ref{sec:twistedindexfromJT}, and indeed $r=1$ is related to the integer charge quantization of the boundary $U(1)_R$ of the 3d $\mathcal{N}=2$ SCFT \cite{Choi:2020baw}.

Combining our computation of the on-shell action \eqref{FE_exp_mix} with \eqref{eq:conjectureZtwisted} and using the appropriate 1-loop determinants of \eqref{FE_JT_with_theta_s6}, we arrive at a gravity prediction for the partition function which includes the low temperature quantum corrections from the Schwarzian mode: 
\begin{equation} \label{ZJT3}
Z_{[H_3;\Sigma_g]}(\beta, \alpha) = e^{ i \alpha \vartheta_*} \sum_{n \in \mathbb{Z}} e^{i  \vartheta_* n} \left( \frac{2 \cos(\pi(\alpha + n))}{\pi(1-4(\alpha +n)^2)} \right) e^{S_* + \frac{2\ \pi^2}{\beta M_{gap}} (1-4 (\alpha+n)^2)} \,,
\end{equation}
We have omitted the details of the derivation, which follow a very similar set of steps to those in \cite{Heydeman:2020hhw,Boruch:2022tno} once one uses the results of Appendix~\ref{app:dimreduction}. Independent of the details (in which we keep only the leading corrections in $S_*$ and $\beta$), in the end we have an effective gravity theory in which the parameters $S_*$, $M_{gap}$, $\vartheta_*$ depend on the details of $H_3$ and $\Sigma_g$ through \eqref{eq:Minimal_StaticDyon_Entropy}, \eqref{eq:vartheta11d}, and \eqref{eq:nearBPSDefs}. In this truncation of the full gravity problem to the effective theory, \eqref{ZJT3} appears as the 1-loop exact answer in this model. In principle, it is also possible to include further $\log N$ corrections by evaluating the 1-loop determinants of the 11d Kaluza-Klein fields around the (near-)BPS metric along the lines of  \cite{Liu:2017vbl}, but this does not affect the leading $\log T$ behavior we study here.

In contrast to the Kerr-Newman case of \cite{Boruch:2022tno,Heydeman:2024ezi}, we now see that the value of $\vartheta$ is nonzero and this will affect the density of states. A crucial point is the large $N$ value of $\vartheta$ presented in \eqref{eq:vartheta11d} as well as the discussion that follows. Naively, $\vartheta \sim \vartheta + 2\pi$ for the path integral to be well defined, and this is required in both the 4d and AdS$_2$ gravity theories, even though the leading large $N$ value of $\vartheta$ \eqref{eq:vartheta11d} does not have the property of being an integer multiple of $\pi$.

Actually, a closer inspection of \eqref{ZJT3} suggests that this periodicity in $\vartheta_* \sim \vartheta_* + 2\pi$ is spoiled by the overall $\alpha$ dependent phase factor. We already understood this from the 4d/3d perspective in \eqref{eq:Z3dcounterterm}, and here we see the manifestation of this from the point of view of super JT gravity. More generally, a nice explanation of this phenomenon is given in \cite{Gaiotto:2017yup}, where the presence of a background connection (here, $\alpha$) means that the system is not invariant under $2\pi$ shifts, but instead only invariant after changing a counter term for this background field. Furthermore, this effect is closely related to the presence of the mixed `t Hooft anomaly in this system, which we describe later in Section \ref{ssec:mixtHooft}.

For generic $\vartheta_*$, the spectrum of the low temperature theory (including both the BPS and non-BPS states) is obtained via Laplace transform, where we use:
\begin{equation} 
Z_{[H_3;\Sigma_g]}(\beta, \alpha) = \int dE \, e^{-\beta E} \rho(\alpha ,E, \vartheta) \, .
\end{equation}
This transform requires several steps that are similar to those in \cite{Stanford:2017thb,Mertens:2017mtv,Boruch:2022tno,Heydeman:2024ezi}, which involves manipulation of $Z(\beta, \alpha)$ by means of Poisson resummation, and a Laplace transform. Because our quantum system (the $\mathcal{N}=2$ Schwarzian mode appearing in the throat) preserves $1d$ $\mathcal{N}=2$ \emph{super-Poincar\'e} symmetry (not related to the broken conformal symmetry), it is natural to write the result $\rho(\alpha ,E, \vartheta)$ as a \emph{density of supermultiplets}. We will sometimes abuse terminology and refer to this as a density of states, but the $\alpha$ dependence of the density of supermultiplets can always be written in terms of (short and long) characters of the $\mathcal{N}=2$ super-Poincar\'e. 

Following the steps outlined in the references above, one obtains
\begin{align} \label{eq:densitytwisted}
\rho (\alpha, E, \vartheta_*) = e^{S_*+i \alpha \vartheta_*} &\left(   \sum_{Z_{Sch} \in \mathbb{Z}-\frac{\vartheta_*}{2\pi}, \, |Z_{Sch}|<\frac12} \! \! \cos(\pi Z_{Sch}) e^{2\pi i \alpha Z_{Sch}} \right. \nonumber \\
&\left. \, \, \, \, \,  + \sum_{Z_{Sch} \in \mathbb{Z}-\frac{\vartheta_*}{2\pi}} (e^{2\pi i \alpha Z_{Sch}} + e^{2 \pi i \alpha (Z_{Sch}-1)})  \frac{\sinh \left( 2 \pi \sqrt{\frac{2 (E-E_{gap})}{M_{gap}}} \right)}{2 \pi E} \Theta(E-E_{gap})\right)\,.
\end{align}
The first line of this formula is the contribution from the BPS states. This is a sum over short multiplets whose charges are only allowed to fall within a window around zero Schwarzian charge. The second line represents the presence of non-BPS long multiplets with charges $(Z_{Sch}, Z_{Sch}-1)$. The spectrum of these states is continuous in the semiclassical approximation, and for a given fixed charge there is generically an energy gap where the continuum appears. This gap depends on the charge and is given by
\begin{equation}
E_{gap} = \frac{M_{gap}}{8} \left(Z_{Sch} -\frac12 \right)^2 \, ,
\end{equation}
with $M_{gap}$ given in \eqref{eq:nearBPSDefs}. A nonzero $\vartheta_*$ angle affects the gap of each charge sector quadratically. The entire density of states is proportional to $e^{S_*}e^{i \alpha \vartheta_*}$, where the first factor sets the extremal degeneracy and the overall scale of the distribution, and the second factor coupling to $\alpha$ can be understood as a shift of the background charge of all states due to the $\vartheta_*$ angle. We next discuss this shift in detail.

\subsection{Dyonic black holes and the Witten Effect}
 This shifting of the background charge really followed from the semiclassical evaluation of the action in \eqref{FE_exp_mix}. It can be understood as a gravity manifestation of the Witten Effect\cite{Witten:1979ey} which occurs for dyons in four dimensional quantum field theories with a $\vartheta$ angle. In general, this effect implies the electric charge of a dyon is no longer necessarily an integer,
\begin{align}
    Z_{\textrm{dyon}} \in \mathbb{Z}-\frac{\vartheta}{2\pi} \, .
\end{align}
This formula was derived in \cite{Witten:1979ey} for dyons in nonabelian gauge theories, but here we are dealing with AdS$_4$ black holes with a $\vartheta$ angle for the graviphoton and we see the same combination appearing in \eqref{eq:densitytwisted}. Naively, the classical electric charge of the extremal black hole is zero (the integer part above is zero) if we want to avoid a naked singularity from \eqref{eq:twistedharmonicfunction}. However, in the quantum theory we work not at fixed electric charge but at fixed potential, and the quantization of the Schwarzian modes tells us there are states whose R-charge has arbitrary integral part. Given these comments, it is convenient to absorb the overall phase of \eqref{eq:densitytwisted} into the Schwarzian charge via:
\begin{align} 
\label{eq:twistedDOS}
\rho (\alpha, E, \vartheta_*) = e^{S_*} &\left(   \sum_{Z_{Sch} \in \mathbb{Z}, \, |Z_{Sch}-\frac{\vartheta_*}{2\pi}|<\frac12} \! \! \cos(\frac{\vartheta_*}{2}) (-1)^{Z_{Sch}}e^{2\pi i \alpha Z_{Sch}} \right. \nonumber \\
&\left. \, \, \, \, \,  + \sum_{Z_{Sch} \in \mathbb{Z}} (e^{2\pi i \alpha Z_{Sch}} + e^{2 \pi i \alpha (Z_{Sch}-1)})  \frac{\sinh \left( 2 \pi \sqrt{\frac{2 (E-\tilde{E}_{gap})}{M_{gap}}} \right)}{2 \pi E} \Theta(E-\tilde{E}_{gap})\right)\,.
\end{align}
Here, we have redefined the Schwarzian charge to absorb the factors of $\vartheta_*$. The second line appears to be independent of this parameter, but the redefinition of the charge results in a new gap scale:
\begin{equation}
\label{eq:twistednewgapscale}
\tilde{E}_{gap} = \frac{M_{gap}}{8} \left(Z_{Sch} -\frac12 -\frac{\vartheta_*}{2\pi} \right)^2 \, .
\end{equation}

We have arrived at this result \eqref{eq:twistedDOS} based on the low-temperature analysis of the bulk black hole solution, quantizing the supergravity modes in the AdS$_2$ throat region using $\mathcal{N}=2$ JT supergravity as a sort of low temperature effective field theory. However, by the AdS/CFT correspondence, we are really making a prediction for the near-extremal behavior of the strongly coupled dual 3d $\mathcal{N}=2$ SCFT of Class $R$ arising from a large number of $N$ M5 branes wrapping $H_3$ in the presence of magnetic flux. The advantage of our approach is that this density of states contains information about both the microcanonical index (which can be compared directly with a field theory calculation performed using the 3d-3d correspondence), but also the physical degeneracies of both the BPS and non-BPS states. These degeneracies are much more difficult to compute directly in field theory, and are expected to be well approximated by random matrix theory with 2 supercharges \cite{Stanford:2019vob,Turiaci:2023jfa,Johnson:2023ofr,Johnson:2024tgg}.

These results are expected to hold quite generally (for other compactifications leading to AdS$_4$ dyonic black holes), but our specific example for 3d $\mathcal{N}=2$ Class $R$ theories explicitly has the additional parameter $\vartheta$, leading to an interesting dependence of the spectrum on $N$. Recalling again that the four dimensional $\vartheta$ angle may be written as
\begin{align}
\label{eq:varthetaN3}
    \vartheta = cs(H_3) \frac{2N^3-N}{3} + \mathcal{O}(N) \, ,
\end{align}
with $cs(H_3)$~\eqref{eq:CSH3} already $2\pi$ periodic, the desired total $2\pi$ periodicity in $\vartheta$ is not achieved since $\frac{2N^3-N}{3}$ is not generically an integer. As mentioned previously, it is believed and argued in \cite{Genolini:2021qbi} that subleading corrections will cure this problem. The spectrum itself is almost\footnote{We have already introduced the idea that the partition function may fail to be periodic in the presence of background fields due to an anomaly\cite{Gaiotto:2017yup}. In this example, this happens when $\alpha \neq 0$, and we describe the anomaly in more detail below.} periodic in $\vartheta \rightarrow \vartheta + 2\pi$, as can be seen by studying \eqref{eq:twistedDOS}. 

 For most of the discussion we will work with generic $\vartheta*$, but the most interesting statements about the quantum spectrum rely on this special condition ($\vartheta_*$ is an integer multiple of $\pi$) which we will eventually implicitly assume. Among other things, when $\vartheta_* =0$ and $\vartheta_* = \pi$ (mod $2\pi$), the theory has charge conjugation invariance\cite{Turiaci:2023jfa}, a remnant of the $CP$ or $T$ invariance in higher dimensions. Given our earlier reported value of \eqref{eq:vartheta11d} which is determined by the hyperbolic Chern-Simons invariant of $H_3$ \eqref{eq:CSH3} (which is periodic), it is understood that $\vartheta$ mod $2\pi$ can (in a coarse grained sense) take any value $0 \leq \vartheta \leq 2\pi$. This is based on ``experimental'' evidence in which $\vol(H_3)$ and $cs(H_3)$ may be computed numerically for a large class of hyperbolic knot complements\cite{knotinfo}. Throughout this work, we do not make specific statements about any particular $H_3$, but manifolds which give values of $\vartheta$ exactly or approximately odd multiple of $\pi$ are the most interesting. There seem to be a number of interesting candidates in \cite{knotinfo}, but this is not an exhaustive list.

\subsection{Dependence of the index and degeneracy on $\vartheta$}
 \label{sec:twistedindexfromJT}
 We have given a prediction for the density of supermultiplets in \eqref{eq:twistedDOS}, and we would like to understand the microcanonical spectrum of the theory. We will conclude this section with our result, summarized in Figure \ref{Fig4}. Before doing this however, we first make a brief detour to explain how our result for the near-BPS partition function is consistent with the known microscopic index.
 
 The sudden appearance of \eqref{ZJT3} based on our arguments from the bulk may have seemed ad-hoc to some readers. As reviewed in the introduction, the physics arguments leading to this proposal pass through a particular near-BPS limit, a truncation of the full gravity problem, an expansion in large area and small temperature, and finally an exact quantization of the classical theory outlined in Appendix~\ref{app:dimreduction}. These approximations are controlled in our expansion, but we do not claim an exact formula for unprotected states. 
 
However, a strong consistency check of our proposal for near-BPS magnetic AdS$_4$ black holes is the comparison between the direct supersymmetric limit of our answer \eqref{ZJT3} and the exact large $N$ topologically twisted index of 3d $\mathcal{N}=2$ Class R Theories computed via the 3d-3d correspondence\cite{Gang:2019uay,Choi:2020baw,Genolini:2021qbi}\footnote{See also \cite{Benini:2019dyp,Bobev:2019zmz} on the computation of the superconformal index in the same framework.}. As we will see, the correct limit to take in the grand canonical ensemble is:
\begin{equation} 
\mathcal{I}_{[H_3;\Sigma_g]} \equiv  Z_{[H_3;\Sigma_g]}(\beta, \alpha = \frac12) \, .
\end{equation}
Intuitively, setting $\alpha = \frac12$ is equivalent to inserting $(-1)^F$ due to the bose/fermi charge quantization of the microscopic theory, and the Schwarzian fermions are rendered periodic in time\footnote{A slightly more general half integral value of $\alpha$ leads to the same conclusion.}. As usual in supersymmetric index computations, periodic fermions may cancel bosonic superpartners, and this can be seen in \eqref{eq:twistedDOS}, where setting $\alpha = \frac12$ results in a cancellation between the bosons and fermions found in long multiplets in the second line. 

From the path integral, JT gravity has supplied the 1-loop determinants which have an interesting behavior in the supersymmetric limit,
\begin{align}
    \lim_{\alpha \rightarrow \frac12} \left( \frac{2 \cos(\pi(\alpha + n))}{\pi(1-4(\alpha +n)^2)} \right) = \begin{cases}\frac{1}{2}, & n=-1,0 \\ 0, & n \neq-1,0 \, .\end{cases} 
\end{align}
Applying this to the terms of the sum in \eqref{ZJT3}, most of the saddles give a vanishing contribution. The supersymmetric limit contains only two saddles, and this was true with or without the nontrivial $\vartheta_*$ angle; the difference is that this additional phase reduces the index from the value obtained from the area alone. Adding the saddles that remain gives:
\begin{equation} 
\label{eq:Indexfieldtheory}
\mathcal{I}_{[H_3;\Sigma_g]} = \frac12 \left ( e^{S_* + i \frac{\vartheta}{2}}+e^{S_* - i \frac{\vartheta_*}{2}}\right ) = e^{S_*}\cos\left(\frac{\vartheta_*}{2}\right)\,,
\end{equation}
This reproduces the large $N$ microscopic result of \cite{Choi:2020baw,Genolini:2021qbi}. Note that in these proposals, the phase of the index is still difficult to determine using the 3d/3d correspondence alone. There, it was still necessary to use some bulk input to argue for this phase, but certain analytically continued solutions were used. Our method using JT gravity with supersymmetric boundary conditions indeed selects out a special set of saddle points, and the normalization leads to the factor of $\frac12$ from the 1-loop determinants. Strictly speaking, none of these works fully fix the prefactor for the saddles, which may receive additional logarithmic corrections\footnote{In this regard we point out that various results on the log-corrections to the entropy of supersymmetric extremal twisted black holes appeared in \cite{Liu:2017vbl,Liu:2018bac,Gang:2019uay,PandoZayas:2020iqr,Bobev:2023dwx}. Because these corrections are independent of the expansion parameter $\beta$, we could incorporate them by shifting our definition of $S_*$.}.

We now discuss the interesting special cases of this formula, as well as the closely related \eqref{eq:twistedDOS} in the supersymmetric limit.

\subsubsection*{Twisted index ($\alpha=\frac12)$ for $\vartheta = 0$ (mod $2\pi$)}
In the case where $\vartheta_*=0$, we have no anomaly and can compare the protected index with the large $N$ degeneracy. We have explained that $\alpha = 1/2$ leads to the Bose-Fermi cancellation in the second line of \eqref{eq:twistedDOS}.  In more detail, the two exponential terms that depend on $\alpha$ are the contribution from the two superpartners in a (continuous) set of long multiplets together, and they cancel when $\alpha = \frac12$. The resulting topologically twisted index only receives contributions from the first line. On the other hand, one can pass to the microcanonical \emph{degeneracy} as opposed to the index by Fourier transform of the first line of \eqref{eq:twistedDOS} and selecting the value $Z_{sch}=0$ allowed by the BPS bound. The degeneracy is exponential in the area $S_*$ and there are no cancellations.

To best explain this result and, in anticipation of the case when $\vartheta_* \neq 0$, we recall that the structure of the Hilbert space $\mathcal{H}$ for $\mathcal{N}=2$ JT gravity is\cite{Turiaci:2023jfa}:
\begin{align}
    \mathcal{H} = \mathcal{H}_\textrm{short} \oplus \mathcal{H}_\textrm{long}\, ,
\end{align}
where we could additionally grade this space by the total R-charge. This is already the structure present in \eqref{eq:twistedDOS}, where $\mathcal{H}_\textrm{short}$ comprises exactly degenerate BPS states at $E=0$, and $\mathcal{H}_\textrm{long}$ form a continuum of near BPS states whose approximate degeneracies depend on $E$. In reference to the ground state degeneracies, we make the distinction
\begin{align}
\label{eq:indexvsdegeneracy}
    \textrm{Index} \equiv \textrm{Tr}_{\mathcal{H}}(-1)^F \, , \quad \textrm{Degeneracy} \equiv \textrm{Tr}_{\mathcal{H}_\textrm{short}} + \textrm{Tr}_{\, \lim_{E\rightarrow 0} \mathcal{H}_\textrm{long}}\, ,
\end{align}
For a non-supersymmetric extremal black hole, the semiclassical Hawking answer gives a finite result for the last term above, equal to the exponential of the extremal entropy. In the present supersymmetric case, we observe that: 
\begin{align}
\label{eq:Index=Deg}
    \textrm{Index}|_{\vartheta_* = 0} = \textrm{Degeneracy}|_{\vartheta_* = 0} = \exp(S_{BH})\, .
\end{align}
This fact has often been used to justify the use of the index as a microscopic proxy for the macroscopic black hole entropy.

Actually, the case where $\vartheta_*$ is an even multiple of $\pi$ reveals the index is only $4\pi$ periodic, a subtle difference from the case where $\vartheta_*$ is absent. In any event, the absolute value of the index agrees with the usual area law degeneracy.

\subsubsection*{Anomalous twisted index ($\alpha=\frac12)$ for $\vartheta = \pi$ (mod $2\pi$)}
For a more general $\vartheta_*$ approaching $\vartheta_* =\pi$ (mod $2\pi$),  the index is reduced and vanishes at the time reversal anomaly point, which was already a possibility given the field theory answer \eqref{eq:Indexfieldtheory}. The advantage of our JT derivation is that one can explicitly see that this vanishing is not directly due to supersymmetric cancellations owing to a $(-1)^F$, as the first line of \eqref{eq:twistedDOS} has already isolated the short multiplets which may not cancel. Instead, the anomalous $\vartheta_*$ parameter results in an exponential reduction in the number of BPS states. Alternatively, the condition:
\begin{align}
    |Z_{Sch}-\frac{\vartheta_*}{2\pi}|<\frac12 \, , \quad Z_{Sch} \in \mathbb{Z} \, ,
\end{align}
represents the allowed BPS bound and charge quantization in $\mathcal{N}=2$ JT with the inequality strict (equality would only be possible  for $\vartheta_*$ an exact odd multiple of $\pi$, but the first line of \eqref{eq:twistedDOS} vanishes there anyway). We do not actually claim to have an example $H_3$ which gives exactly $\vartheta_* = \pi$, but we have already noted that the numerical evidence in \cite{knotinfo} suggests one may be able to get close. 

At this point, it is worth noting that one can find a different reason for the index to vanish. This is based on a mixed `t Hooft anomaly present in this problem, which we show in Section \ref{ssec:mixtHooft}. This anomaly appears because the classical symmetries $\alpha \rightarrow \alpha + 1$ and $\alpha \rightarrow -\alpha$ do not always leave the partition function invariant. The failure of these symmetries obstructs the index being nonzero when the anomaly is present.

 For now, we emphasize that in this anomalous case, there can still be some notion of ground state degeneracy coming from the long multiplets, but their interpretation is more subtle and will be discussed shortly.

\subsubsection*{Near BPS degeneracy for $0 \leq \vartheta < \pi$ and ABJM}
In the situation where we have no anomaly, the non-BPS states are gapped, according to \eqref{eq:twistednewgapscale}. The microscopic ground state degeneracy is $e^{S_*}$, in agreement with the Hawking area law \eqref{eq:Index=Deg}. The spectrum of non-BPS multiplets (appearing above the gap scale at order $\sim N^{-3}$) is qualitatively similar to that found in \cite{Heydeman:2022lse} for $\mathcal{N}=4$ SYM. Our arguments would suggest a similar spectrum for the 3d ABJM theory compactified on $S^1 \times \Sigma_g$ which has a vanishing $\vartheta$ angle, but otherwise identical dyonic black hole solutions in the minimal 4d $\mathcal{N}=2$ gauged supergravity truncation. This is implied by our discussion in Sec.~\ref{Sec:dyonicsolutions} combined with the result of \cite{Genolini:2021qbi} which shows $\vartheta=0$ for M2 branes and $SE_7$ compactifications.

In the M5/Class $R$ case where $\vartheta_*$ can be nontrivial, we can contemplate the effect of increasing $\vartheta_*$. As the value of $\vartheta_*$ is increased from $0$ to $\pi$, the $Z_{sch}=1$ multiplet gap begins to close as a consequence of \eqref{eq:twistednewgapscale}. Further, the ground state degeneracy in the first line of \eqref{eq:twistedDOS} decreases, as we already saw in our discussion of the index. 

\begin{figure}[H]
	\begin{center}
	\includegraphics[width=0.67\linewidth]{"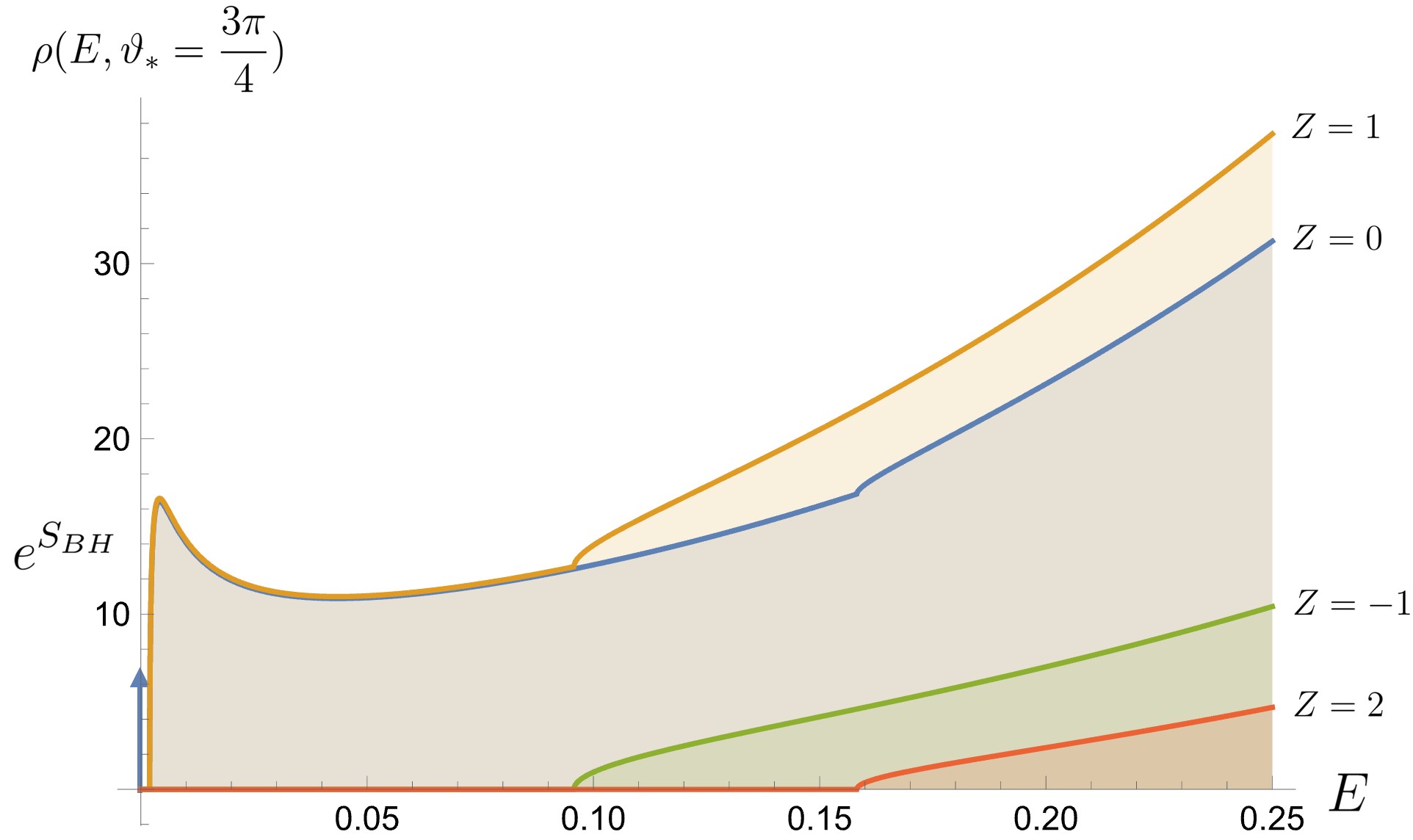"}
 \end{center}
	\caption{A plot of the BPS and non-BPS spectral density for an assortment of charge sectors labeled by integers $Z$. We have set $M_{gap}=1$ and $\vartheta_* = \frac{3\pi}{4}$ for concreteness. The $Z=0$ (blue curve) admits BPS states at zero energy and a bump in the non-BPS spectrum at the gap scale of another $Z=0$ state in a different multiplet. The mass gap for $Z=0$ and $Z=1$ close as $\vartheta_*$ approaches $\pi$. The $Z=-1$ and $Z=2$ sectors are always gapped and do not permit BPS states. As $\vartheta_*$ approaches $\pi$, the $Z=0$, $Z=1$ curves merge, similarly the $Z=-1$, $Z=2$ merge. This doubling of the spectrum is a hallmark of the anomalous $CT$ symmetry.} \label{Fig4}
	\end{figure}

A simple Fourier transform in $\alpha$ of \eqref{eq:twistedDOS} allows one to extract the microcanonical density of states $\rho\left(E, Z_{sch}, \vartheta_* \right)$. We do this in more detail in the next section, but here plot the result for generic $0 \leq \vartheta_* < \pi$ in Figure \ref{Fig4}. We can examine the BPS and near-BPS spectrum for different fixed charge sectors, where each curve is a sum of two families of states (with different respective gap scales). These microcanonical densities exhibit the same closing of the gap as the supermultiplet densities do for certain charge sectors.

On these grounds, perhaps our most interesting results concern the spectrum of non-BPS states when $\vartheta_*$ is exactly an odd multiple of $\pi$, which deserves its own subsection below.

\subsection{Near-BPS states at $\vartheta = \pi$ (mod $2\pi$)}
\label{ssec:anomalousdegeneracy}
At exactly $\vartheta_* = \pi$, the BPS degeneracy is zero (in the gravity approximation that lead to \eqref{eq:twistedDOS}), and the only states that survive are long multiplets. We again emphasize that this vanishing is not due to cancellations, but instead a large $N$ quantum supergravity effect\footnote{The approximations in this paper do not exclude all BPS states in the microscopic field theory, and there may still exist such states; the only statement is that their degeneracy is subleading compared to $\mathcal{O}(\exp(N^3))$.}. Additionally, the $Z_{sch}=1$ supermultiplets are ungapped, and there is a divergence in their spectrum at $E=0$, similar to the $\mathcal{N}=1$ Schwarzian theory \cite{Stanford:2017thb}. This means that the final term in \eqref{eq:indexvsdegeneracy} is not completely well defined, and it involves a limit which we will discuss around \eqref{eq:threashholdDOS}.  The total density of supermultiplets (at fixed potential) is the weighted infinite sum of all the constituent curves for different values of $Z_{sch}$. 

While we oriented the above discussion for $\vartheta_* = \pi$ specifically, the same spectrum is obtained for $\vartheta_*$ equal to any odd integer multiple of $\pi$, so below we take
\begin{align}
\vartheta_* = (2k - 1) \pi \, , \quad k \in \mathbb{Z} \, .
\end{align}
There is always a corresponding special value of $Z_{sch}$, which we call $Z_{sch}^* = k$, that is gapless and typically of order $N^3$ due to the 11d formula \eqref{eq:varthetaN3}. We would now like to focus on these gapless states in the microcanonical ensemble.

\begin{figure}
	\begin{center}
	\includegraphics[width=0.65\linewidth]{"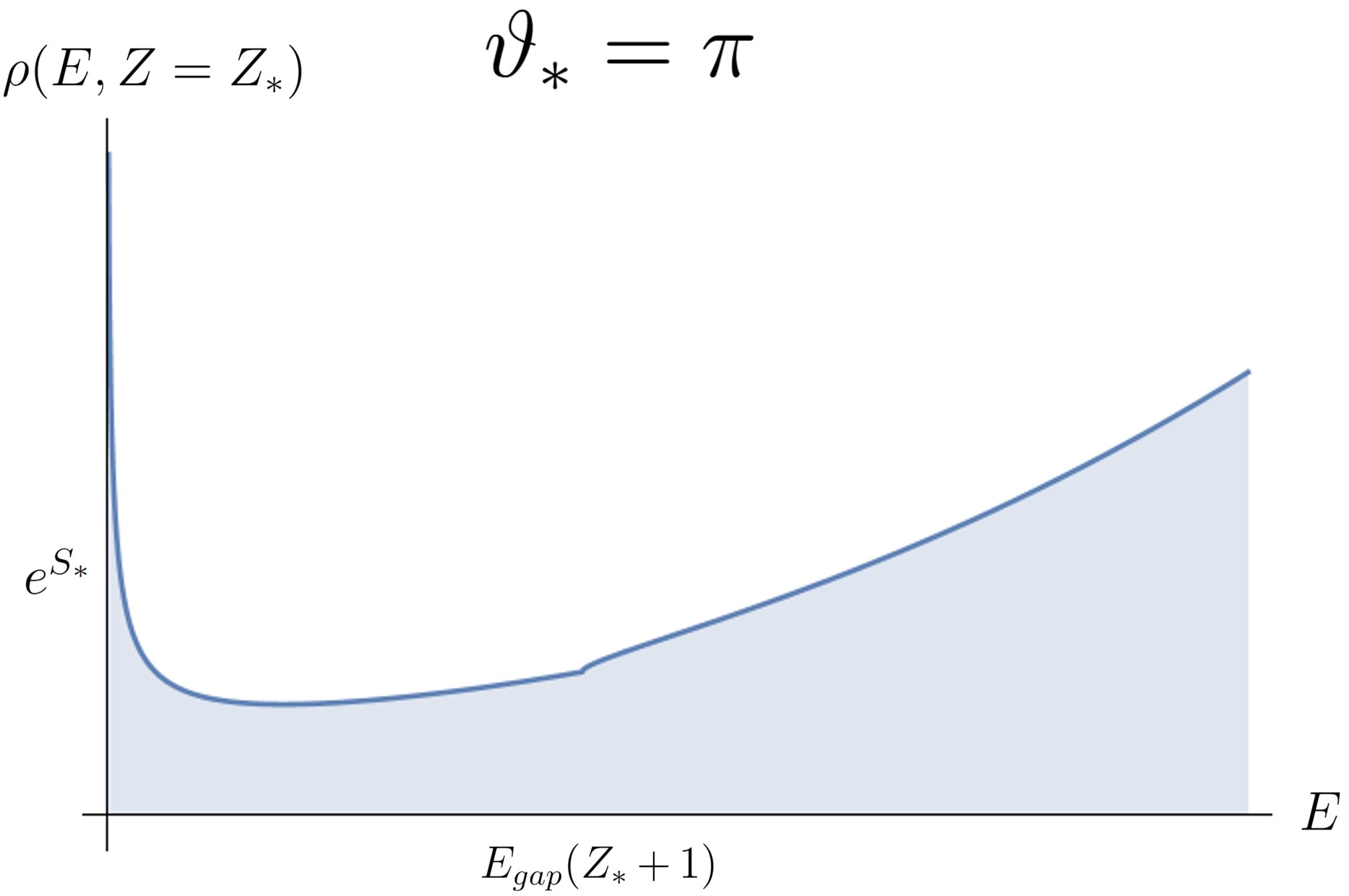"}
 \end{center}
	\caption{The plot displays the density of states for $\vartheta_* = \pi $. One can see that the density of states diverges at the origin due to a single gapless set of states with charges $Z_{sch}=Z_*$, where $Z_*$ depends on $\vartheta_*$ such that $E_{gap}(Z_*) =0$. We expect this divergence is a consequence of the gravity approximation and not a fundamental aspect of the theory; more on this is explained at the end of this section. The remaining states are gapped and begin contributing at fixed nonzero energies, resulting in the bump in the curve at $E_{gap}(Z_*+1)$.} \label{Fig3}
	\end{figure}

Performing a simple Fourier transform of \eqref{eq:twistedDOS} in $\alpha$, the microcanonical distribution for a fixed value of $Z_{sch}$ is:
\begin{align} 
\label{eq:twistedMICRO}
\rho\left(E, Z_{sch}, \vartheta_* = (2k - 1) \pi\right) &= \frac{e^{S_*}}{2 \pi E}  \sinh \left( 2 \pi \sqrt{\frac{2 (E-\frac{M_{gap}}{8} \left(Z_{Sch} -k \right)^2)}{M_{gap}}} \right) \Theta(E-\frac{M_{gap}}{8} \left(Z_{Sch}-k  \right)^2)  \nonumber \\
& + \left(Z_{sch} \longleftrightarrow Z_{sch}+1 \right) \, , 
\end{align}
where we indicated that this spectrum is a sum of 2 identical terms with $Z_{sch}$ replaced by $Z_{sch}+1$. These states actually have the same charge, but come from different supermultiplets in \eqref{eq:twistedDOS}.
Therefore, this distribution genuinely counts states of fixed charge, rather than multiplets labeled the the charge of the top component. The gap indeed vanishes for $Z_{sch} = k$. We plot this density in Fig.~\ref{Fig3}, specifically choosing $\vartheta_* = \pi$, $Z_{sch}=Z^*_{sch}=1$. The main visible effect comes from the first line of \eqref{eq:twistedMICRO}, but there is a bump in the curve coming from the contribution of states from the ``adjacent'' supermultiplet, which enter at their respective gap energy. 

We will discuss some consequences of this formula below, but for now we note that the microcanonical degeneracies actually grow at sufficiently low energies when we fix the charge to be that of a gapless multiplet. For these charges, the degeneracies are large, coming as a multiple of the semiclassical $e^{S_*}$ answer, but there are no BPS states and the index vanishes.

In the language of the decomposition \eqref{eq:indexvsdegeneracy}, we thus are lead to the following surprising conclusion: 
\begin{align}
\label{eq:indexdegeneracytheta}
    \textrm{Index}|_{\vartheta_* = \pi}\, \, \neq \,\,\textrm{Degeneracy}|_{\vartheta_* = \pi} \,\,\neq \,\,\exp(S_{BH}) \, .
\end{align}
Of course, it was not guaranteed that all of these quantities were supposed to agree once quantum corrections are taken into account. 

We have already commented on the apparent divergence in the spectrum for the gapless states, which we claim are the origin of the large mismatch in \eqref{eq:indexdegeneracytheta}. This requires further discussion, as it makes it difficult to define the ``ground state degeneracy'' which we claim does not really agree with the Hawking formula. For this special set of gapless states which have $Z_{sch} = k$, we may examine the $E\rightarrow 0$ behavior of \eqref{eq:twistedMICRO}:
\begin{align} 
\label{eq:inversesqrtedge}
\rho\left( E\rightarrow 0, Z_{sch}=k, \vartheta_* = (2k - 1) \pi \right) = e^{S_*}\sqrt{\frac{2}{M_{gap}}}\frac{1}{\sqrt{E}} \biggr ( 1 + \mathcal{O}(M_{gap}^{-1}E ) \biggr) \, .
\end{align}
This is a sign that there is a number of (near) ground states which grows exponentially in $N$, but with an inverse square root of the energy-- all with a fixed value of the R-charge. This might have been the best we could hope to do using only the gravitational theory, and even with quantum corrections these states cannot be exactly resolved in gravity. If one could do this (with additional non-perturbative and stringy effects), it is the counting of these states which would deserve to be called the extremal degeneracy in \eqref{eq:indexdegeneracytheta}.

To understand the growth of \eqref{eq:inversesqrtedge}, we can restore the dependence on $N$ using \eqref{eq:nearBPSDefs}:
\begin{align} 
\rho \left(E\rightarrow 0 \right ) \approx \sqrt{\left(\frac{S_*}{\sqrt{2} \pi E} \right)} \, e^{S_*} =  \sqrt{\left(\frac{(g-1)\vol(H_3)}{3\sqrt{2} \pi^2 E} \right)} N^{3/2}\, e^{\frac{(g-1)\vol(H_3)}{3\pi}N^3}\, ,
\label{eq:threashholdDOS}
\end{align}
where the expansion relied on $E$ being smaller than the gap scale set by $N^3$ in 11-dimensional units (note that there is no gap, but there is a gap scale $M_{gap}$). It is worth commenting that the $N^{\frac32}$ prefactor may compete with other logarithmic corrections we have not included which go beyond the Schwarzian limit. We can further estimate the total number of states in an energy window by integrating over energies up to $N^{-3}$. We find
\begin{equation}
    \int_0^{\frac{M_{gap}}{8}}\! \! \!  dE \, \, \rho \, (E \rightarrow 0) = e^{S_*} \, ,
\end{equation}
using a judicious choice of the upper limit of integration. This tells us that the approximate number of black hole microstates is still $\propto \exp(\frac{\textrm{Area}}{4})$ for the extremal area, but they have exponentially small energy spacings.

As an alternative to the gravity problem, it would be very interesting to understand this set of non-BPS states from the point of view of the 3d superconformal field theory. However, one would seem to need to do a strongly coupled CFT calculation to find these\footnote{Despite the current lack of a microscopic field theory derivation, it is possible to recast the higher dimensional SCFT near extremality in terms of more solvable supersymmetric SYK-like models\cite{Benini:2022bwa,Heydeman:2022lse,Benini:2024cpf}. For $\mathcal{N}=4$ SYM at weak coupling, there have also been some attempts to see the Schwarzian behavior in  \cite{Chang:2023zqk,Cabo-Bizet:2024gny}.}. It is unclear if the usual rules of the 3d-3d correspondence is helpful in this situation, as we have neither supersymmetry nor a weakly coupled limit. Optimistically, there exist simpler theories which realize the $\vartheta_* = (2k-1)\pi$ anomaly but also have a weakly coupled description.

Despite our inability to compute the this anomalous spectrum directly starting from the 3d Class $R$ theories, there is good evidence to believe the kind of spectrum displayed in Figure~\ref{Fig3} may arise from an ordinary quantum mechanical theory. The $\mathcal{N}=2$ SYK model\cite{Fu:2016vas} may be regarded as a toy model of the strongly coupled field theory problem, and it is amenable to direct analytical and numerical techniques. If one considers this model with an odd $N$ number of fermions, one may obtain a spectrum as in \cite{Stanford:2017thb}, Fig.~(1). In fact, the supersymmetric SYK model with an even number of fermions results in the $\vartheta_* =0$ answer, while an odd number gives our current example. Since the Hilbert space is finite dimensional, there is no true divergence at $E=0$, and one may numerically find the ground state energies to be exponentially small in $N$\cite{Heydeman:2022lse}\footnote{A similar spectrum may be obtained by starting with $\mathcal{N}=2$ SYK and softly breaking supersymmetry\cite{Heydeman:2024ohc}. The result has an exponentially large number of non-BPS states with exponentially small energies.}, consistent with their gravity expectation that these states are resolved by non-perturbative effects. What is surprising here is that we find the same behavior displayed for a black hole with an 11d uplift. Given the relation \eqref{eq:varthetaN3}, there is some reason to believe the integrality of $N$ plays a similar role for the 3d Class $R$ theories as it does for SYK, where $N$ is now the number of M5 branes.

In the comments above, we have tried to argue that the apparent divergence of the density of states at $E=0$ is an artifact of the gravity calculation, and the expectation is that a more refined computation (either in the dual field theory or in gravity) would resolve the spectrum and there would exist a state, possibly with degeneracy, of lowest energy. A more intuitive picture of the divergence which can be made very concrete was given in \cite{Johnson:2024tgg}. This relies on the understanding that the density of states obtained at the disc approximation in supergravity is the first term of the expansion of a particular double scaled random matrix model. In this random matrix context, eigenvalues tend to repel\footnote{Unless there is a symmetry that makes them degenerate. When a macroscopic number of BPS states exists, the repulsion is macroscopic and results in a gap.} and distribute across the energy spectrum subject to some potential. Because our theory has unbroken supersymmetry, there is a hard wall at $E=0$ where eigenvalues may not penetrate. However, due to the mixed 't Hooft anomaly, there are actually no BPS states in this theory which have $E=0$ exactly; essentially we have a BPS bound, but this bound is never saturated at large $N$. This means there is no eigenvalue repulsion from $E=0$, but high energy states exert a force on lower energy states, with no balancing repulsion from the BPS sector. The energy must remain positive, but an arbitrarily large number of eigenvalues pile up near $E=0$. The specific $E^{-\frac12}$ shape of the edge is the one consistent with the random matrix interpretation\cite{Stanford:2019vob,Turiaci:2023jfa}.

\subsubsection*{Speculations on the fate of the ground state}
To summarize the above, when $\vartheta_* = (2k-1)\pi$, the leading gravity calculation for $Z_{sch}=k$ charges leads to an exotic low energy spectrum. The true fate of the system at exponentially small energies is beyond the regime of the gravity approximation. We cannot exclude the possibility that there exist exact BPS states at $E=0$ for these Class $R$ theories, but the consistency of the gravity picture requires that if these states exist, their degeneracies are exponentially suppressed in $N$ relative to the non-BPS states.

It is worth mentioning the possibility that there may exist no $H_3$ and choice of $N$ which realizes $\vartheta$ an exact \emph{odd} multiple of $\pi$. There seem to many examples close to this in \cite{knotinfo}, and there are explicitly manifolds with a vanishing $\vartheta$. A more detailed analysis is outside the scope of our current paper, but it may help to resolve these issues about the ground state.

Additionally, one interesting clue is that this amplification of the number of states at low energies occurs only in a particular R-charge sector which depends on $\vartheta_*$, so this sector may be interesting to study even from a pure conformal field theory viewpoint. Along these lines, it would be interesting if one could use the existence of the anomaly at to make more concrete predictions about the ground states similar to \cite{Gaiotto:2017yup}. In that work, it was argued that 4d gauge theory with $\vartheta = \pi$ has a mixed `t Hooft anomaly between center symmetry and time reversal, and this anomaly ensures the ground state is twofold degenerate. A hint of this kind of degeneracy in our problem comes from taking \eqref{eq:twistedMICRO} for different values of $Z_{sch}$ together. Since each value of the charge comes with a sum of two terms, there are actually two different sets of multiplets which become gapless when we consider all charge sectors. This is at least suggestive that the true ground state has this twofold degeneracy.

To strengthen this argument and the analogy with \cite{Gaiotto:2017yup}, we turn our attention to demonstrating this mixed anomaly `t Hooft anomaly explicitly in our gravity calculation.

 \subsection{The mixed `t Hooft anomaly at $\vartheta = \pi$}
 \label{ssec:mixtHooft}

 Our proposal for the low temperature partition function and spectrum of magnetic AdS$_4$ black holes is that which matches a particular version of $\mathcal{N}=2$ JT gravity, or equivalently the $\mathcal{N}=2$ super-Schwarzian theory. The parameters of this effective theory are determined using properties of the UV, in particular the classical higher dimensional solution (as well as mild input from the existence of a dual boundary CFT$_3$ with quantized charges). This is to be expected, as the Schwarzian theory is an effective field theory describing the infrared/near horizon physics; in particular it is largely determined by the pattern of preserved and broken symmetries. Up to numerical parameters such as the extremal entropy and gap scale/coupling constant, the JT answer is largely independent of the choice of compactification. Beyond the classical gravity analysis on this background, we would like to treat the Schwarzian theory as an honest quantum theory, and it may be that certain classical symmetries are actually anomalous. The topological term does not affect the classical equations of motion, but we have shown the resulting quantum density of states is dramatically changed. When $\vartheta = \pi$ (mod $2\pi$), we would like to explain this phenomenon in part due to the presence of a mixed 't Hooft anomaly which is visible at finite temperature.

The classical 2d/1d theory has at the boundary an unbroken $U(1)_R$ symmetry as well as an unbroken $C$ symmetry which acts by charge conjugating fermions and reversing the sign of the charge\cite{Turiaci:2023jfa}\footnote{One could potentially have both $C$ or $CT$ discrete symmetries, and here $C$ seems to be the most natural starting with the $CP$ violating term in four dimensions. The time reversal anomaly in 4d thus manifests as a $C$ anomaly in 2d.}. However, the enlargement of $U(1)_R \times \mathbb{Z}_2$ to $O(2)$ in the quantum theory relies on these symmetries being non-anomalous. At finite temperature with a background connection $\alpha$ for the $U(1)_R$ turned on, this anomaly can be diagnosed by checking whether the partition function is still invariant under the classical symmetries\cite{SeibergPITP}.

As we will show below, our result for the partition function can only be made invariant by one but not both of these symmetries, implying that there is a mixed anomaly. An anomaly of this kind has interesting implications for four-dimensional gauge theories\cite{Gaiotto:2017yup} with a theta angle, and perhaps the simplest system which has this general kind of mixed 't Hooft anomaly is that of the quantum mechanics of a free Dirac fermion, described in the appendix of that work and reviewed in \ref{App:anomaly}. As in this simple example, we may act on the partition function \eqref{ZJT3} with the action of the $\alpha \rightarrow -\alpha$ symmetry which implements the $\mathbb{Z}_2$ charge conjugation, as well as the $\alpha \rightarrow \alpha + 1$, which sets the charge quantization of $U(1)_R$. We observe, for general $\vartheta_*$,
\begin{align}
    Z_{[H_3;\Sigma_g]}(\beta, \alpha+1) &=  e^{ i \alpha \vartheta_* + i \vartheta_*} \sum_{n \in \mathbb{Z}} e^{i  \vartheta_* n} \left( \frac{2 \cos(\pi(\alpha + 1 + n))}{\pi(1-4(\alpha +1 +n)^2)} \right) e^{S_* + \frac{2\ \pi^2}{\beta M_{gap}} (1-4 (\alpha+1+n)^2)} \nonumber \, \\
    &= Z_{[H_3;\Sigma_g]}(\beta, \alpha)\, .
\end{align}
The invariance of the partition function under $\alpha \rightarrow \alpha + 1$ was essentially guaranteed from the completion $\alpha \rightarrow \alpha + n$ in the original classical action \eqref{FE_exp_mix} and the sum over these saddles. This is somewhat different than the results of \cite{Heydeman:2022lse,Turiaci:2023jfa} due to the additional classical $i\alpha \vartheta_*$ term in the action.

Under the time reversal symmetry, we have
\begin{align}
\label{eq:timereverseZ}
    Z_{[H_3;\Sigma_g]}(\beta, -\alpha) &=  e^{ -i \alpha \vartheta_*} \sum_{n \in \mathbb{Z}} e^{-i  \vartheta_* n} \left( \frac{2 \cos(\pi(\alpha  + n))}{\pi(1-4(\alpha +n)^2)} \right) e^{S_* + \frac{2\ \pi^2}{\beta M_{gap}} (1-4 (\alpha+n)^2)} \nonumber \, \\
    &= \overline{Z}_{[H_3;\Sigma_g]}(\beta, \alpha)\, .
\end{align}
As expected, this acts by complex conjugation. We see that for generic $\vartheta_*$, the charge conjugation is always anomalous, even for $\vartheta_* = \pi$ (mod $2\pi$) due to the prefactor. This means that the partition function which was natural starting with the sum over classical solutions has a charge conjugation anomaly when we turn on a nontrivial background holonomy $\alpha$ for $U(1)_R$. This establishes the mixed anomaly between the $U(1)_R$ and $\mathbb{Z}_2$ symmetries.

One can contemplate a more general version of the $\mathcal{N}=2$ Schwarzian partition function in which the values of $\vartheta_*$ appearing in the classical prefactor $i \alpha \tilde{\vartheta}_*$ and the sum over saddles $i \vartheta_* n$ are not the same; $\vartheta_* \neq \tilde{\vartheta}_*$. This is unnatural from the point of view of \eqref{FE_exp_mix} but is more similar to the version of this theory studied in \cite{Turiaci:2023jfa}. In this case, the $U(1)_R$ symmetry is anomalous for any $\vartheta_* \neq \tilde{\vartheta}_*$, but charge conjugation is restored for $\tilde{\vartheta}_*=0$ and $\vartheta_* = \pi$.

In Section \ref{sec:twistedindexfromJT}, we studied the grand canonical index and found from the explicit formula that it vanishes when $\vartheta_* = (2k-1)\pi$, $k \in \mathbb{Z}$. We can also understand this from the more abstract point of view of the anomaly, essentially via a contradiction argument. Recall from that section that the index is computed in the JT language from $\alpha=\frac12$, or more generally an odd multiple of $\frac12$, which ensures cancellation of the long multiplets in \eqref{eq:twistedDOS}. Since $\alpha = \frac12$ can be fixed by the combined set of $\alpha \rightarrow \alpha+1$ and $\alpha \rightarrow -\alpha$, but the partition function is not in general invariant under both these transformations, we will run into a problem. We can first combine the two facts we learned about the behavior of the full partition function for generic $\alpha$ and $\vartheta_*$:
\begin{align}
    Z_{[H_3;\Sigma_g]}(\beta, \alpha+1, \vartheta_*) = Z_{[H_3;\Sigma_g]}(\beta, \alpha, \vartheta_*) \, , \qquad Z_{[H_3;\Sigma_g]}(\beta, -\alpha,\vartheta_*) = \overline{Z}_{[H_3;\Sigma_g]}(\beta, \alpha,\vartheta_*) \, ,
\end{align}
which implies when $\alpha = \frac12$ that,
\begin{align}
\label{eq:realindex}
    Z_{[H_3;\Sigma_g]}(\beta, \alpha = \frac12 , \vartheta_*) = \overline{Z}_{[H_3;\Sigma_g]}(\beta, \alpha = \frac12,\vartheta_*) \equiv \mathcal{I}_{[H_3;\Sigma_g]}(\vartheta_*) \, .
\end{align}
Now, on  the other hand we could take generic $\alpha$, but fixed $\vartheta_* = (2k-1)\pi$. The infinite sum over saddles for both $Z$ and its complex conjugate are the same in \eqref{eq:timereverseZ}, so we can take the ratio:
\begin{align}
    \frac{Z_{[H_3;\Sigma_g]}(\beta, \alpha, \vartheta_* = (2k-1)\pi)}{\overline{Z}_{[H_3;\Sigma_g]}(\beta, \alpha,\vartheta_* = (2k-1)\pi)} = e^{2\pi i \alpha (2k-1)} \, .
\end{align}
Upon setting $\alpha=\frac12$, we see these partition functions satisfy $Z = - Z$, potentially in contradiction with \eqref{eq:realindex}. This is avoided if $\mathcal{I}_{[H_3;\Sigma_g]}(\vartheta_* = (2k-1)\pi))$ simply vanishes for integral $k$\footnote{Strictly speaking, we were not supposed to take the ratio when the function vanishes, but the main conclusion about extracting the overall phase leads to the same result.}. When $k$ is instead half-integral, there is no anomaly and the partition functions are simply equal, allowing for a non-vanishing index.

We end this section with one further comment which was outside the scope of our analysis. One hallmark of the mixed anomaly is that it may be ``moved around'' by the inclusion of counterterms and redefinitions of the charge operator. This ambiguity means there is a presentation of the partition function in which charge conjugation is restored, but the $U(1)_R$ is improperly quantized and thus anomalous. At least when $\vartheta_* = \pi$, we expect to see this by interpreting the Schwarzian density of states as a trace and shifting the definition of the microscopic charge operator by a half integer. Thus, one way to interpret the $\exp(i \alpha \vartheta_*)$ phase is as a shift of the background charge. It would be interesting to derive this version of the partition function starting from the four-dimensional gravitational path integral.

\section{Discussion}
\label{sec:discussion}
In this work, we have explored the effect of topological terms on the spectrum of near-BPS black holes. In quantum field theory, such terms will generally impact the low energy theory and have microscopic effects for the ground states and the spectrum of charges. In gravity, one is typically doing semi-classical calculations which capture coarse grained information about the spectrum, and in this situation it may seem reasonable to ignore topological effects. However, we have demonstrated that the $\vartheta$-angle anomaly in tandem with the near-extremal quantum effects may lead to very different conclusions, even for coarse-grained observables. This realizes a possibility first pointed out in the context of JT gravity and SYK in \cite{Stanford:2017thb}, and here we have a precise M-theory realization.

The Witten effect for dyonic black holes described in this work essentially moves an exponentially large number of would-be BPS states outside of the BPS bound dictated by $\mathcal{N}=2$ JT gravity. For the non-BPS states, we find a full continuous family of states becomes gapless as $\vartheta_*$ approaches $\pi$. Thus the anomaly obstructs the existence of BPS states, but supersymmetry still guarantees the energy must remain positive. The only conclusion is that in the leading $N$ approximation, an exponentially large number of states accumulate near $E=0$. The true fate of the ground state would appear to require information outside of this gravity approximation, and we expect the ground state energy to be exponentially small in $N$. Further, the kinds of arguments presented in \cite{Gaiotto:2017yup} could optimistically suggest the ground state is two-fold degenerate. In any event, a key conclusion of our work is that there exist (near) extremal black holes with a large area whose entropy is not accurately reproduced by either the Bekenstein-Hawking formula nor the supersymmetric index.

Our gravity computation suggests that (for certain values of $N$), the low energy limit of 3d SCFTs of Class $R$ is highly nontrivial. This phase of the field theory may be difficult to analyze directly, and it is unclear if one may use the 3d/3d correspondence to study this regime. If a similar phase is present in more conventional Lagrangian SCFTs, a starting point may be models similar to those in \cite{Benini:2022bwa}. 

Finally, we make the observation that a gravitational path integral calculation which does not take into account topological terms and anomalies will not generally be valid. For familiar contributions such as the 4d $\vartheta$-angle, this may be straightforward. But there can be more subtle discrete phases which are not 
 obvious to identify. We do not know of a general classification of which topological field theories are non-trivial in the presence of a black hole, but we expect this to be an important piece of information in any microscopic quantum gravity calculation.

\section*{Acknowledgments}

We thank P. Benetti Genolini, N. Bobev, J. Boruch, A. Castro, K. Hristov, L. Iliesiu, J. Maldacena, S. Murthy, L. Pando Zayas, M. Rangamani, V. Reys, D. Stanford, A. Strominger, C. Vafa, O. Varela, E. Witten for helpful comments, interesting discussions, and collaborations on related topics. We are grateful to Z. Komargodski and G. Turiaci for a careful reading of the draft, and for useful comments. MTH is supported by Harvard University and the Black Hole Initiative, funded in part by the Gordon and Betty Moore Foundation (Grant 8273.01) and the John Templeton Foundation (Grant 62286). MTH completed part of this work while at the Kavli Institute for Theoretical Physics (KITP), supported in part by grant NSF PHY-2309135. The work of CT is supported by the Marie Sklodowska-Curie Global Fellowship (Horizon 2020 Program) SPINBHMICRO-101024314. This work was performed in part at Aspen Center for Physics, which is supported by National Science Foundation grant PHY-2210452. CT also acknowledges support from the Simons Center for
Geometry and Physics, Stony Brook University where some of this research was performed.

\begin{appendix}
\section{Reduction of the (bosonic) theory to 2d}
\label{app:dimreduction}

We will follow the same strategy and conventions as \cite{Castro:2021wzn}, where the 2d reduction of more general supergravity theories was performed\footnote{See also \cite{Castro:2022cuo} for the cosmological Einstein Maxwell theory without the $\vartheta$ term.}. The four dimensional bosonic action is
\begin{equation}\label{eq:4daction}
I_{4d}=-{1\over 16\pi G} \int d^4x \sqrt{g}\left(R-{2\Lambda}- F_{\mu\nu}F^{\mu\nu}\right) \, .
\end{equation}
We will do a dimensional reduction of this theory to two dimensions, where we take the following ansatz for the metric and field strength
\begin{equation}\label{eq:metric4d}
ds^2_4=g_{\mu\nu} dx^\mu dx^\nu= {\Phi_0\over \Phi} g^{(2)}_{ab} dx^a dx^b + \Phi^2 d\Omega_2^2~, \qquad F= F_{ab} dx^a \wedge dx^b + P \sinh \theta d\theta \wedge d\phi ~.
\end{equation}
In this setup, we have introduced a scalar dilaton field $\Phi(x)$, depending only on  the 2d coordinates $x^a$, $a,b=(t,r)$. Here $g^{(2)}_{ab}$ is the two-dimensional metric, which similarly is assumed to depend only on $(t,r)$. The constant $\Phi_0$ introduced below is the value of the dilaton field in the IR solution, which corresponds to the unperturbed near horizon geometry $AdS_2 \times \Sigma_g$.  The parameter $P$ is the magnetic charge. The gauge field $F_{\mu\nu}$ is taken to be dyonic.
The effective 2d action that one obtains is (we have denoted with $R^{(2)}$ the Ricci scalar of the 2d metric $g^{(2)}_{ab}$)
\begin{equation}\label{eq:2daction}
I_{2d}=-{(g-1)\over 4 G} \int d^2x \sqrt{g^{(2)}}\Phi^2\left(R^{(2)} -{\Phi\over \Phi_0} F_{ab}F^{ab} + 2{\Phi_0\over\Phi^3}-{2\Lambda}{\Phi_0\over \Phi}- \frac{\Phi_0}{\Phi^5} P^2 \right)~.
\end{equation}
 Notice that this theory is a consistent truncation of \eqref{eq:4daction}.  The  Maxwell's equations can be straightforwardly integrated, and give
\be
F_{ab}= Q {\Phi_0\over \Phi^3}\sqrt{g^{(2)}}\epsilon_{ab}~,
\ee
with parameter $Q$ related to the electric charge. We also record $F_{ab}F^{ab} = -2Q^2 \frac{\Phi_0^2}{\Phi^6}$. If we now include a $\vartheta$ term, namely we start from an action
\begin{equation}\label{eq:4daction_theta}
I_{4d}=-{1\over 16\pi G}  \int d^4x \sqrt{g}\, \left(R-{2\Lambda}- F_{\mu\nu}F^{\mu\nu}\right) + \frac{i\vartheta}{8\pi^2} \int F \wedge F  \,,
\end{equation}
we obtain an effective 2d action
\begin{eqnarray}\label{eq:2daction_theta}
I_{\rm 2d} &= &- \frac{(g-1)}{ 4 G} \int d^2x \sqrt{-g}\Phi^2\left(R^{(2)} -{\Phi\over \Phi_0} F_{ab}F^{ab}+ 2{\Phi_0\over\Phi^3}-{2\Lambda}{\Phi_0\over \Phi} - \frac{\Phi_0}{\Phi^5} P^2 \right) \nonumber \\
&+ & \frac{i (g-1) \vartheta P}{2 \pi} \int dx^a dx^b F_{ab} .
\end{eqnarray} 
Variation of the action with respect to the metric and the dilaton gives respectively (the $\vartheta$ term is topological)
\be\label{eq:eom2}
(\nabla_a\nabla_b -g^{(2)}_{ab}\square) \Phi^2 +g^{(2)}_{ab}\left({1\over 2}{\Phi^3\over \Phi_0} F^2 -{\Phi_0\over \Phi^3} P^2 +{\Phi_0\over \Phi} -\Lambda {\Phi_0 \Phi} \right)=0~,
\ee
and
\be\label{eq:eom1}
R^{(2)}  -{3\over 2}{\Phi\over \Phi_0} F^2 +\frac{3P^2 \Phi_0}{ \Phi^5} -  {\Phi_0\over \Phi^3} -\Lambda {\Phi_0\over \Phi} =0 ~.
\ee

The IR background we want to expand around is the $AdS_2 \times \Sigma_g$ near horizon geometry, and it is characterized by a constant dilaton, $\Phi_0$. Imposing this, from \eqref{eq:eom2} and \eqref{eq:eom1} we find
\be \label{eq:ads2}
Q^2 +P^2 = \Phi_0^2(1-\Lambda \Phi_0^2) ~,
\qquad 
R^{(2)}_0 = - \frac{2}{\Phi_0^2}  \left(1-2\Lambda \Phi_0^2\right) \,,
\ee
setting the AdS$_2$ radius of the locally AdS$_2$ background space $\bar g^{(2)}_{ab}$ to be
$
\frac{1}{\ell_2^2}= \frac{1}{\Phi_0^2}  \left(1-2\Lambda \Phi_0^2\right). 
$ Now we can consider linear fluctuations around this background solution. We define
\begin{align}\label{eq:lin3}
\Phi &= \Phi_0 + \epsilon \mathcal{Y}(x)~,\cr
g_{ab}&= \bar g^{(2)}_{ab} + \epsilon h_{ab}~,
\end{align}
where $\epsilon$ is the perturbation parameter. We are ready then to find the equations of motion for the fields $\mathcal{Y}(x)$ and $h_{ab}$. From \eqref{eq:eom1} and \eqref{eq:eom2}, expanding at linear order in $\epsilon$,  we obtain
\be\label{eq:lin2}
\bar \nabla^a\bar \nabla^b h_{ab} -\bar \square h(x) + {1\over \ell_2^2} h(x)- {4\over \Phi_0^3}(3-4\Lambda \Phi_0^2) \mathcal{Y}(x)=0 \,,
\ee
where $h(x)= h_{ab}\bar g^{(2) ab}$, and
\begin{align}\label{eq:lin1}
(\bar \nabla_a\bar \nabla_b -\bar g^{(2)}_{ab}\bar \square) \mathcal{Y}(x)+{1\over \ell_2^2}\bar g^{(2)}_{ab} \mathcal{Y}(x) =0 \,.
\end{align}
The latter equation is the equation for the dilaton field in JT gravity. Eq. \eqref{eq:lin2} can be solved 
straightforwardly upon making a choice of gauge for $\bar g^{(2)}_{ab}$ and $h_{ab}$, as done for instance in \cite{Maldacena:2016upp,Castro:2018ffi}.

\section{The $O(2)$ Anomaly}
\label{App:anomaly}
As an example of a simple system which has the same kind of mixed `t Hooft anomaly as $\mathcal{N}=2$ JT with $\vartheta \neq 0$, we consider a supersymmetric extension of the theory of a single qubit discussed in \cite{Gaiotto:2017yup}. Consider the quantum mechanics of a single Dirac fermion and auxiliary complex scalar:
\begin{align}
    L = \bar{\psi}\partial_t \psi - \bar{b} b \, .
\end{align}
The Hamiltonian of this theory vanishes, but it has complex supersymmetries generated by infinitesimal Grassmann $\epsilon$, $\bar{\epsilon}$:
\begin{align}
    \delta_\epsilon \psi = \epsilon \bar{b} \, , \quad \delta_\epsilon b = \epsilon \partial_t \bar{\psi} \, , \quad \delta_{\bar{\epsilon}} \bar{\psi} = \bar{\epsilon} b \, , \quad \delta_{\bar{\epsilon}} \bar{b} = \bar{\epsilon} \partial_t \psi \, ,
\end{align}
which leave the theory invariant up to a total derivative. 

More importantly for this discussion, the theory classically has a $O(2) = U(1) \ltimes \mathbb{Z}_2 $ global symmetry which is the phase rotation of the complex fields and the charge conjugation symmetry, respectively. We can define the fermions $\psi$ and $\bar{\psi}$ to have charges $q$, $-q$. The charges for $b$,$\bar{b}$ are determined by whatever charge we assign $\epsilon$, $\bar{\epsilon}$, which in this simple free model are arbitrary. The transformations are:
\begin{align}
    U(1):& \,\,\, \psi \rightarrow e^{i q} \psi \, , \quad \bar{\psi} \rightarrow e^{-i q} \bar{\psi} \, , \\
    \mathbb{Z}_2:& \,\,\, \psi \rightarrow \bar{\psi} \, , \quad \bar{\psi} \rightarrow \psi \, .
\end{align}
Because the $O(2)$ symmetry does not commute with the supersymmetry transformations, it is the R-symmetry of the model. The only role of the $b$, $\bar{b}$ is to realize supersymmetry, and we are free to integrate out these fields. The Hilbert space of the theory is just that of the free complex fermion, and there are two states. The grand canonical partition function is determined by placing this theory on a thermal circle and introducing a background field for the $U(1)_R$ part of the R-symmetry, where $\alpha = \frac{1}{2\pi}\oint A \sim \alpha + 1/q$ is the holonomy for this background connection. This periodicity condition is fixed by the quantization condition for the charge $q$ we assigned. We then promote the ordinary derivative to a gauge covariant derivative with this connection, and note that again the $b$ fields do not pick up any phase. The partition function is thus
\begin{align}
    Z(\beta, \alpha) = \textrm{Tr} \, e^{-\beta H + 2\pi i \alpha R} = 1 + e^{2\pi i q \alpha} \, . 
\end{align}
This has the manifest $U(1)_R$ periodicity,
\begin{align}
    Z(\beta, \alpha + 1/q) = Z(\beta, \alpha) \, ,
\end{align}
but note that in this language the $\mathbb{Z}_2$ symmetry acts as $\alpha \rightarrow -\alpha$ and we have
\begin{align}
    Z(\beta, -\alpha) \neq Z(\beta, \alpha) \, .
\end{align}
Since the partition function is not invariant under the $\mathbb{Z}_2$ when we turn on a background field for the $U(1)_R$, the $\mathbb{Z}_2$ is anomalous with these choices.

We can attempt to cure the anomaly by changing the charge of the Clifford vacuum. This for instance may be achieved by adding a half-integer $k=\frac12$ quantized 1d Chern-Simons term. Then the partition function is particle/hole symmetric and we have
\begin{align}
    \tilde{Z}(\beta, \alpha) = e^{-i \pi q \alpha} +  e^{i \pi q \alpha} = 2 \cos(\pi q \alpha)\, .
\end{align}
With this choice, the partition function satisfies $\tilde{Z}(\beta, -\alpha) = \tilde{Z}(\beta, \alpha)$, but the periodicity condition $\alpha \sim \alpha + 1/q$ and thus the charge quantization is lost (the gauge field is not periodic). While the classical theory had the full $O(2)$, there is no scheme in which the grand partition function is invariant under this full symmetry, so there is a mixed anomaly between $U(1)_R$ and $\mathbb{Z}_2$. 

Note that in the case where we have $N$ copies of the original theory, the partition function is
\begin{align}
    \tilde{Z}_N(\beta, \alpha) = \left ( 2\cos(\pi q \alpha)\right)^N \, ,
\end{align}
and this function is invariant under both symmetries when $N$ is even. So the anomaly only occurs for odd $N$, and this is analogous to the behavior found in the $\mathcal{N}=2$ Schwarzian theory with $\vartheta =0$ and $\vartheta = \pi$.

This anomaly persists for the (refined) supersymmetric index, which is obtained by setting $\alpha = \frac{1}{q}(\frac{r}{q}-\frac12 )$ where $r$ is a discrete potential; $r=0,1,\dots q$. In this case\cite{Fu:2016vas}, we have
\begin{align}
    I(r) = 1-e^{2\pi i \frac{r}{q}} \, , \\
    \tilde{I}(r) = 2 \sin\left (\frac{\pi r}{q} \right) \, .
\end{align}
In the former expression, the index is not real, and this is reflected in the spectrum which contains a chargeless boson and a charged fermion. In the latter expression, the periodicity in $r\sim r+q$ is lost. In the scheme where we shifted the vacuum charge, the charge spectrum is now symmetric under conjugation, but all charges are fractional.

\end{appendix}

\bibliographystyle{./JHEP-2}
\bibliography{main.bib}

\end{document}